\documentclass{aa}
\usepackage{natbib}
\usepackage{graphicx}
\usepackage{rotating}
\usepackage{times}
\usepackage{amsmath}
\usepackage{txfonts}
\usepackage{multirow}
\usepackage{color}

\usepackage[T1]{fontenc}
\usepackage{aecompl}
\usepackage{amsfonts,amssymb}
\usepackage{scrextend}
\usepackage{textcomp}

\newcommand{\nht}{\ifmmode {{\rm NH}_3} \else {NH{\bas 3}} \fi}
\newcommand{\tco}{\ifmmode {^{13}{\rm CO}} \else {$^{13}{\rm CO}$}\fi}
\newcommand{\dco}{\ifmmode {^{12}{\rm CO}} \else {$^{12}{\rm CO}$}\fi}
\newcommand{\cdo}{\ifmmode {{\rm C}^{18}{\rm O}} \else {${\rm C}^{18}{\rm O}$}\fi}

\newcommand{\htco}{\ifmmode {{\rm H}^{13}{\rm CO}^{+} } \else {${\rm H}^{13}
{\rm CO}^{+}$ }\fi}
\newcommand{\hco}{\ifmmode {{\rm H}^{12}{\rm CO}^{+} } \else {${\rm H}^{12}
{\rm CO}^{+}$ }\fi}
\newcommand{\juz}{\ifmmode {{\rm J}=1\rightarrow 0} \else
{J=1$\rightarrow$0}\fi}
\newcommand{\jdu}{\ifmmode {{\rm J}=2\rightarrow 1} \else
{J=2$\rightarrow$1}\fi}
\newcommand{\jtd}{\ifmmode {{\rm J}=3\rightarrow 2} \else
{J=3$\rightarrow$2} \fi}
\newcommand{\jcq}{\ifmmode {{\rm J}=5\!\rightarrow\!4} \else
{${\rm J}=5\!\rightarrow\!4$} \fi}
\newcommand{\as}{\ifmmode {^{\scriptscriptstyle\prime\prime}}
        \else $^{\scriptscriptstyle\prime\prime}$\fi}
\newcommand{\am}{\ifmmode {^{\scriptscriptstyle\prime}}
        \else $^{\scriptscriptstyle\prime}$\fi}
\newcommand{\hh}{\ifmmode {{\rm H}_2} \else {H$_2$} \fi}
\renewcommand{\hco}{\ifmmode {{\rm HCO}^+} \else {HCO$^+$} \fi}
\newcommand{\hhco}{\ifmmode {{\rm H}_2{\rm CO}} \else {H$_2$CO} \fi}
\newcommand{\ddco}{\ifmmode {{\rm D}_2{\rm CO}} \else {D$_2$CO} \fi}
\newcommand{\chhdoh}{\ifmmode {{\rm CH}_2{\rm DOH}^+} \else {CH$_2$DOH} \fi}
\newcommand{\chhhod}{\ifmmode {{\rm CH}_3{\rm OD}^+} \else {CH$_3$OD} \fi}
\newcommand{\chhhoh}{\ifmmode {{\rm CH}_3{\rm OH}^+} \else {CH$_3$OH} \fi}
\newcommand{\tchhhoh}{\ifmmode {^{13}{\rm CH}_3{\rm OH}^+} \else {$^{13}$CH$_3$OH} \fi}
\newcommand{\dcop}{\ifmmode {{\rm DCO}^+} \else {DCO$^+$} \fi}
\newcommand{\cchh}{\ifmmode {{\rm C}_2{\rm H}_2} \else {C$_2$H$_2$} \fi}

\newcommand{\hcccn}{\ifmmode {{\rm HC}_3{\rm N}} \else {HC$_3$N} \fi}

\begin{document}
\title{Chemistry in Protoplanetary Disks: the gas-phase CO/H$_2$ ratio and the Carbon reservoir}

\author{
 L.~Reboussin \inst{1,2},
 V.~Wakelam \inst{1,2},
 S.~Guilloteau \inst{1,2},
 F.~Hersant \inst{1,2},
 A.~Dutrey \inst{1,2}}

\institute{
Univ. Bordeaux, LAB, UMR 5804, F-33270, Floirac, France
\and{}
CNRS, LAB, UMR 5804, F-33270 Floirac, France}

\authorrunning{Reboussin et al.} %
\titlerunning{Chemistry in Protoplanetary Disks}

\abstract
{The gas mass of protoplanetary disks, and the gas-to-dust ratio,
are two key elements driving the evolution of these disks and the formation
of planetary system.}
{We explore here to what extent CO (or its isotopologues) can be
used as a tracer of gas mass.}
{We use a detailed gas-grain chemical model and study the evolution of the
disk composition, starting from a dense pre-stellar core composition. We
explore a range of disk temperature profiles, cosmic rays ionization
rates, and disk ages for a disk model representative
of T Tauri stars.}
{At the high densities that prevail in disks, we find that,
due to fast reactions on grain surfaces,
CO can be converted to less volatile forms (principally s-CO$_2$, and to a lesser extent s-CH$_4$) instead of being evaporated
over a wide range of temperature. The canonical gas-phase abundance
of 10$^{-4}$ is only reached above about 30 -- 35 K.
The dominant Carbon bearing entity depends on the temperature
structure and age of the disk. The chemical evolution of CO is also sensitive to the
cosmic rays ionization rate. Larger gas phase CO abundances
are found in younger disks. Initial conditions, such as parent cloud age and
density, have a limited impact.}
{This study reveals that CO gas-phase abundance
is heavily dependent on grain surface processes, which remain
very incompletely understood so far. The strong dependence on
dust temperature profile makes CO a poor tracer of the gas-phase content
of disks.}


\maketitle

\section{Introduction}
Measuring the mass of protoplanetary disks remains a challenging
task. H$_2$ being either unobservable in the physical conditions prevailing
throughout most of the disks, or when detected, only sampling peculiar
conditions \citep{Bary2008}, we have to rely on indirect tracers
to estimate the disk masses.

A common practice is based on the dust emission at mm wavelengths,
where the dust opacity is moderate and thus the emission sensitive
to most of the dust mass.
However, the dust emissivity is not
well known: it depends on
dust characteristics, in particular the grain size distribution, and
differential grain growth has been observed in disks
\citep{Guilloteau2011, Perez2012}, invalidating the use
of a unique emissivity coefficient throughout the disk. Despite this
limitation, under the simple assumption of an average emissivity
of order 2 cm$^2$g$^{-1}$, it has been shown that the dust masses
of protoplanetary disks scale with their host stellar mass
\citep{Andrews2013}.

In any case, dust only cannot inform us about the gas content. It is
often assumed that the gas-to-dust ratio is of order 100 to derive
disk masses, but many processes (such as the vertical dust settling
and the radial drift of large
grains) may affect the gas-to-dust ratio.
HD is perhaps the best proxy for H$_2$, and has been detected so far in
TW Hya \citep{Bergin2013}. However, the interpretation is strongly
dependent on the thermal profile of the disk. Furthermore, there is
no longer any sensitive enough telescope to detect it in more distant
protoplanetary disks.

It has been realized in the early studies of disk emission that
the $^{13}$CO lines were much fainter than expected from mm
continuum emission from dust \citep{Beckwith1990, Dutrey1996}.
Part of this discrepancy may be attributable to different disk sizes
in dust and continuum \citep{Pietu2006, Pietu2007, Hughes2009}, but
depletion of CO on dust grains was invoked as the main cause.

Recently, \citet{Williams&Best2014} suggested that, because of its
relatively simple chemistry, CO could be a reliable tracer of dust mass.
They argued that gas-phase CO abundance depends mostly on gas
temperature, and that with a reasonable knowledge of the disk thermal
structure, the fraction of carbon locked-up in gas-phase CO can be
evaluated (10$^{-4}$ where T > 20 K). Apart from the upper layers, which contain little mass,
the thermal structure in disks is dominated by dust
\citep{Chiang&Goldreich1997}, and can be reasonably well constrained
from SED measurements. Exploring a large grid of disk parameters,
\citet{Williams&Best2014} concluded that disk gas masses were in general
rather small, leading to an average gas-to-dust ratio well below the canonical
value of 100 found in the interstellar medium.

However, this finding contradicts
the high mass suggested by the analysis of the HD observations of TW Hya
\citep{Bergin2013}. To resolve the discrepancy between these two
approaches, a very low CO to H$_2$ ratio is required.
Very low apparent CO to dust ratios were also found
by \citet{Chapillon2008} for CQ Tau and MWC 758, two
Herbig Ae stars, as well as for BP Tau \citep{Dutrey2003}, despite
large temperatures (above 30 K) deduced from CO. Under the simple assumption
that CO is mostly in the gas-phase because of the high temperatures,
\citet{Chapillon2010} showed that the lack of detection of C{\small I}
in CQ Tau excludes photodissociation as the sole cause of low CO column
densities, and implies a very low gas-to-dust ratio.

The CO rotational transitions being in general quite optically thick
in disks, using the less abundant isotopologues is required to sample
the gas content.  This introduces a significant complexity
because of the different shielding of the isotopologues against
photodissociation by the UV radiation field. Another added complexity
is the $^{13}$CO fractionation which can occur in luke-warm regions (20-30 K).
These processes have been studied in detail in \citet{Miotello2014}, who
showed that these effect have substantial impact on the derivation
of the CO gas content, and could reduce the discrepancy between the two approaches.

\citet{Favre2013} concluded
that the emission of C$^{18}$O in TW Hya could only be explained assuming
CO has been converted to other hydrocarbons, rather than released from grains.
As an other alternative, \citet{Chapillon2008} suggested that
because large grains can remain cold, CO might remain trapped on such grains
even in relatively warm disks, like those around the Herbig Ae stars
CQ Tau or MWC 758.

Given that other molecules suffer from an even more complex chemistry than
CO and its isotopologues, understanding to what extent CO can be a
tracer of disk masses is important.

In this paper, we re-explore this issue using a detailed
gas-grain chemical network. We concentrate here on some
aspects of chemistry which can affect the (gas-phase) CO abundance
in protoplanetary disks. We first study the impact of
temperature profile and the role of grain surface chemistry
for disks of different ages. Secondly we evaluate the importance
of initial conditions on the time dependent disk chemistry.

Our purpose is not to build complete disk models, but to reveal
the importance of these effects on the chemical evolution of
protoplanetary disks. The paper is organized as follows.
In section 2, we desccribe our chemical model and the physical disk
structure used for this work. In Section 3, we present the effect of
the disk vertical temperature profile on the chemistry of carbon-bearing
species, as well as the impact of the cosmic rays ionization
rate and the age of the disk. We also study the sensitivity of our
model to the initial C/O ratio, the gas density
and the age of the parent molecular cloud. Section 4 contains a
discussion of our work. Finally, in Section 5 we present our
conclusions about this work.

\section{Modeling}
\subsection{Chemistry}
\subsubsection{Model Description}
We used the Nautilus chemical model \citep{Semenov2010, Reboussin2014}
which computes the abundance of species as a function of time in the
gas-phase and at the surface of the grains. The model includes pure
gas-phase chemistry with bimolecular reactions, ionizations and
dissociations by direct impact of cosmic ray particle and UV photons
(interstellar FUV photons and secondary UV photons induced by cosmic
ray/H$_{2}$ interactions). We assume the standard cosmic ray ionization
rate $\zeta_{\text{CR}}$=1.3$\times$10$^{-17}$s$^{-1}$. The dust grains
are represented by spherical particles with a radius of 0.1\,$\mu$m, a
density of 3\,g.cm$^{-3}$, and are made of amorphous olivine. The gas
and dust temperatures are assumed to be the same and we use a
gas-to-dust mass ratio of 100. The gas-grain interactions include
adsorption of gas-phase species, those species can desorb back into the
gas-phase by thermal desorption and non-thermal desorption (via the
energy released by exothermic surface reactions and cosmic ray induced
desorption). The impact of photodesorption is discussed in Sect. 3.2.
The model takes into account grain-surface reactions with diffusion
reactions at the surface of the dust particle via thermal hopping
(we do not take into account diffusion by quantum tunneling in this study).
The energy barrier between two adjacent sites is taken as half of the
binding energy \citep{Garrod&Herbst2006}. Photodissociation on the surface
by UV photons and cosmic ray induced UV field are also considered.

\subsubsection{Cloud Model Parameters}
To obtain the initial chemical composition of the disk, we first compute
the chemical composition of the parent molecular cloud. We run Nautilus
during 10$^{6}$ yr for typical dense cloud conditions as our nominal
cloud: a gas density of
n$_{\text{H}}$$\,$=$\,$n(H)$\,$+$\,$2n(H$_{2}$)$\,$=$\,$2$\times$10$^{4}$$\,$cm$^{-3}$,
a temperature of 10 K (with $T_{\text{gas}}$\,$=$\,$T_{\text{dust}}$),
a visual extinction of 10, a cosmic ray ionization rate of
1.3$\times$10$^{-17}$ s$^{-1}$ and the initial abundances as listed in
Table \ref{table1}. For all elements but He, N, C and O, we have used
the ``low metal'' elemental abundances defined by \citet{Graedel1982}.
For oxygen, two values have been considered: a ``low depletion'' case
which corresponds to C/O = 0.5 in our standard model, and a ``high
depletion'' case which corresponds to C/O = 1.2 considered in
Sect.\,\ref{sec:sub:elemental}.
The helium abundance is assumed to be 0.09 \citep{Wakelam&Herbst2008}
while the nitrogen and carbon abundances are extrapolated from
\citet{Jenkins2009} up to a density of 2$\times$10$^{4}$$\,$cm$^{-3}$
\citep[see discussion in][]{Hincelin2011}. 

\begin{table}[h!]
 \centering
 \begin{minipage}{19cm}
  \caption{Initial abundances used in our model.}
  \begin{tabular}{ll}
  \hline \hline
   Element & Abundance relative to H\\
\hline
H$_{2}$&0.5\\
He&{9$\times10^{-2}$
   \footnote{See discussion in \citet{Wakelam&Herbst2008}.}}\\
N&{6.2$\times10^{-5}$
 \footnote{\label{bb}\citet{Jenkins2009}.}}\\
O&{3.3$\times10^{-4}$
  \footnote{\label{cc}See discussion in \citet{Hincelin2011}.}}\hspace{1cm}(C/O=0.5)\\
&{1.4$\times10^{-4}$
  \footref{cc}}\hspace{1cm}(C/O=1.2)\\
C$^{+}$&{1.7$\times10^{-4}$
   \footref{bb}}\\
S$^{+}$&{8$\times10^{-9}$
    \footnote{\label{dd}Low metal abundances \citep{Graedel1982}.}}\\
Si$^{+}$&{8$\times10^{-9}$
   \footref{dd}}\\
Fe$^{+}$&{3$\times10^{-9}$
   \footref{dd}}\\
Na$^{+}$&{2$\times10^{-9}$
   \footref{dd}}\\
Mg$^{+}$&{7$\times10^{-9}$
   \footref{dd}}\\
P$^{+}$&{2$\times10^{-10}$
   \footref{dd}}\\
Cl$^{+}$&{1$\times10^{-9}$
   \footref{dd}}\\
\hline
\label{table1}
\end{tabular}
\end{minipage}
\end{table}

\subsubsection{Chemical Network}
The chemical network used for this work contains 8624 reactions:
6844 are pure gas-phase reactions and 1780 are grain-surface and
gas-grain interactions. The model follows the chemistry of 703 species
(atoms, radicals, ions and molecules): 504 are gas-phase species and
199 are species on grains. The surface network is based on
\citet{Garrod2007} whereas the gas-phase network is based on
kida.uva.2011 with updates from \citet{Wakelam2013} and
\citet{Loison2014b, Loison2014a}. We have added 19 gas-phase species and
377 pure gas-phase reactions (mainly carbon and nitrogen chains). The
network is available on the KIDA (KInetic Database for Astrochemistry)
website \footnote{http://kida.obs.u-bordeaux1.fr/models}.

\subsection{Disk Structure}

\begin{figure}[h!]
\centering
\includegraphics[scale=0.35]{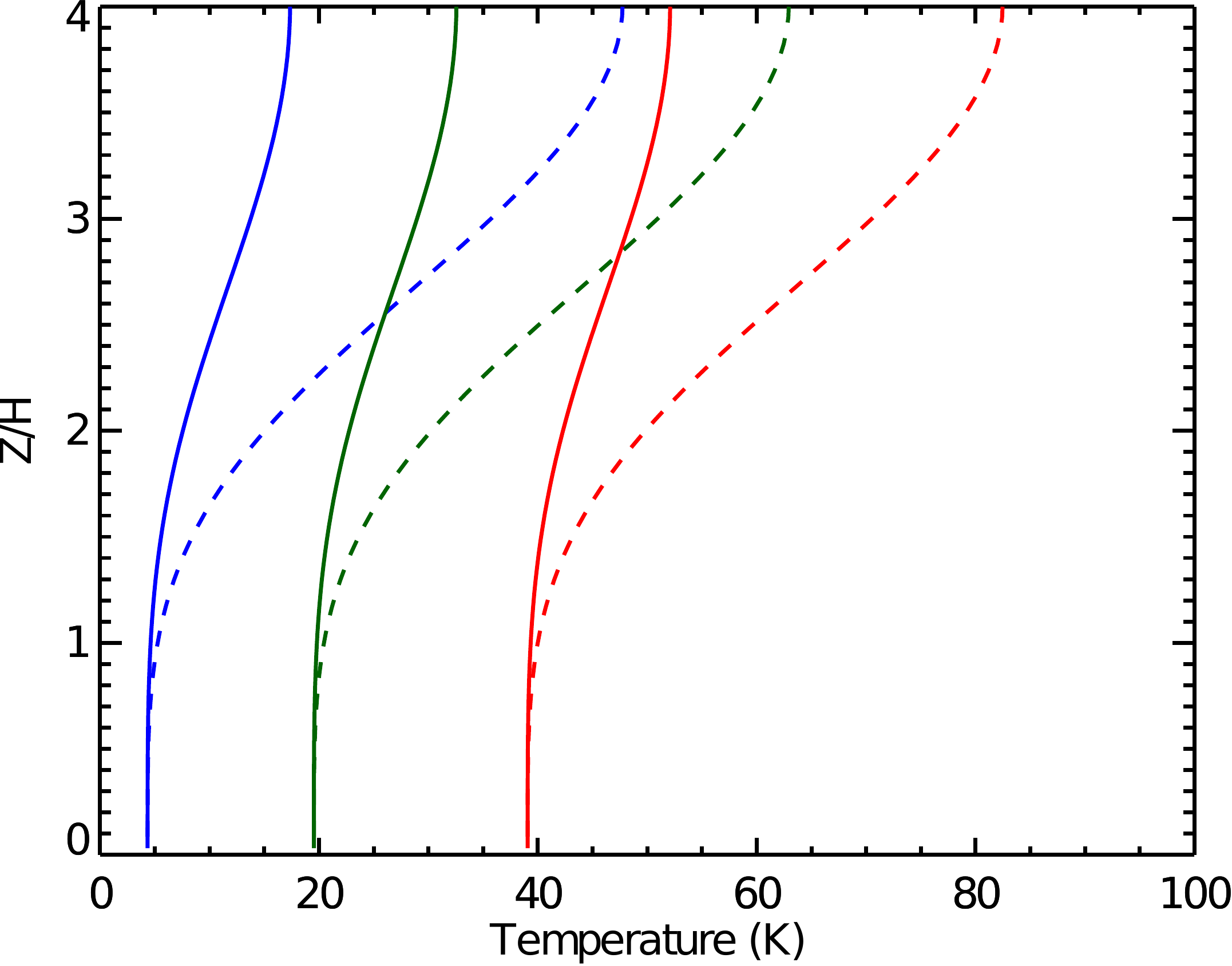}
\vskip2em
\includegraphics[scale=0.35]{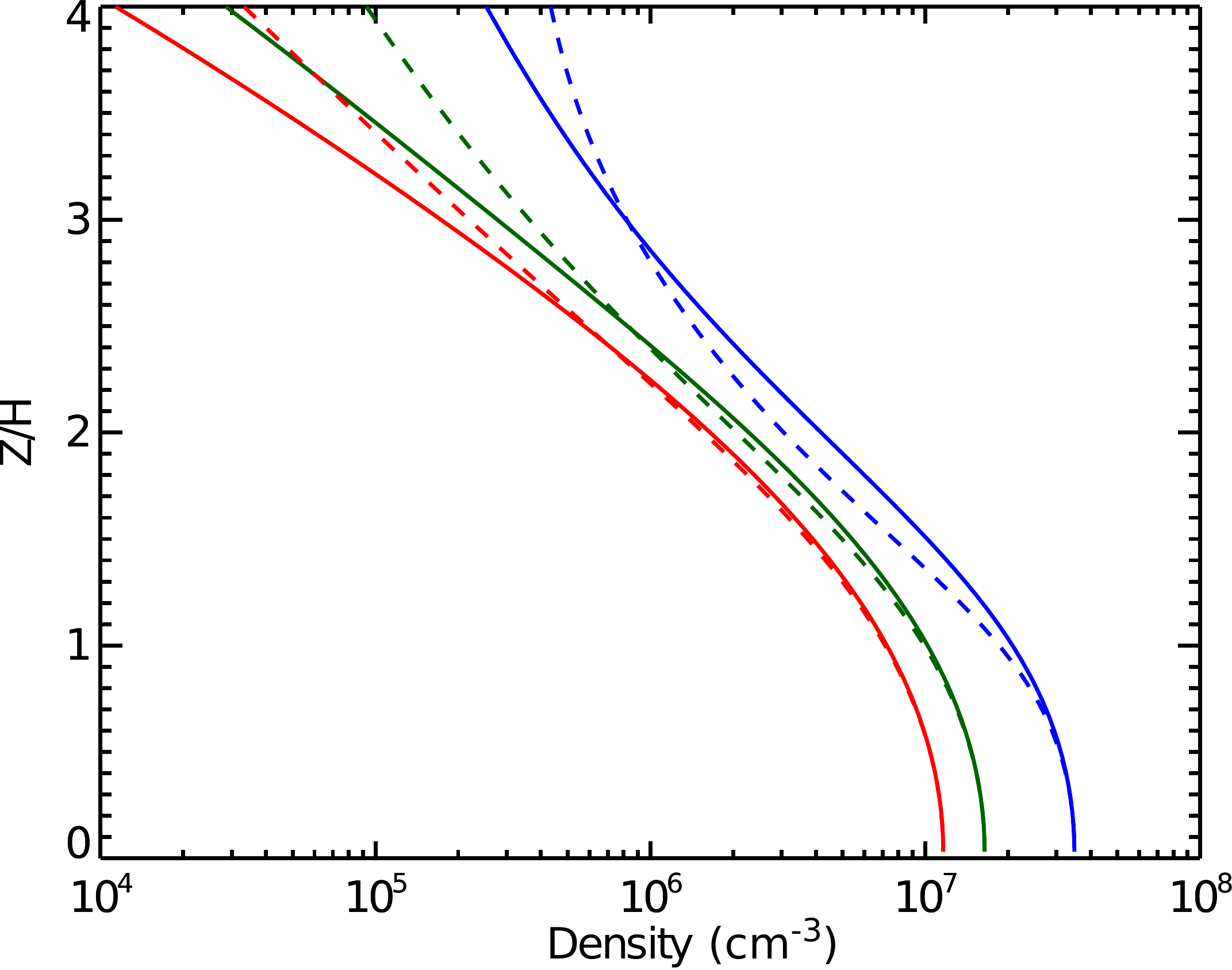}
\vskip2em
\includegraphics[scale=0.35]{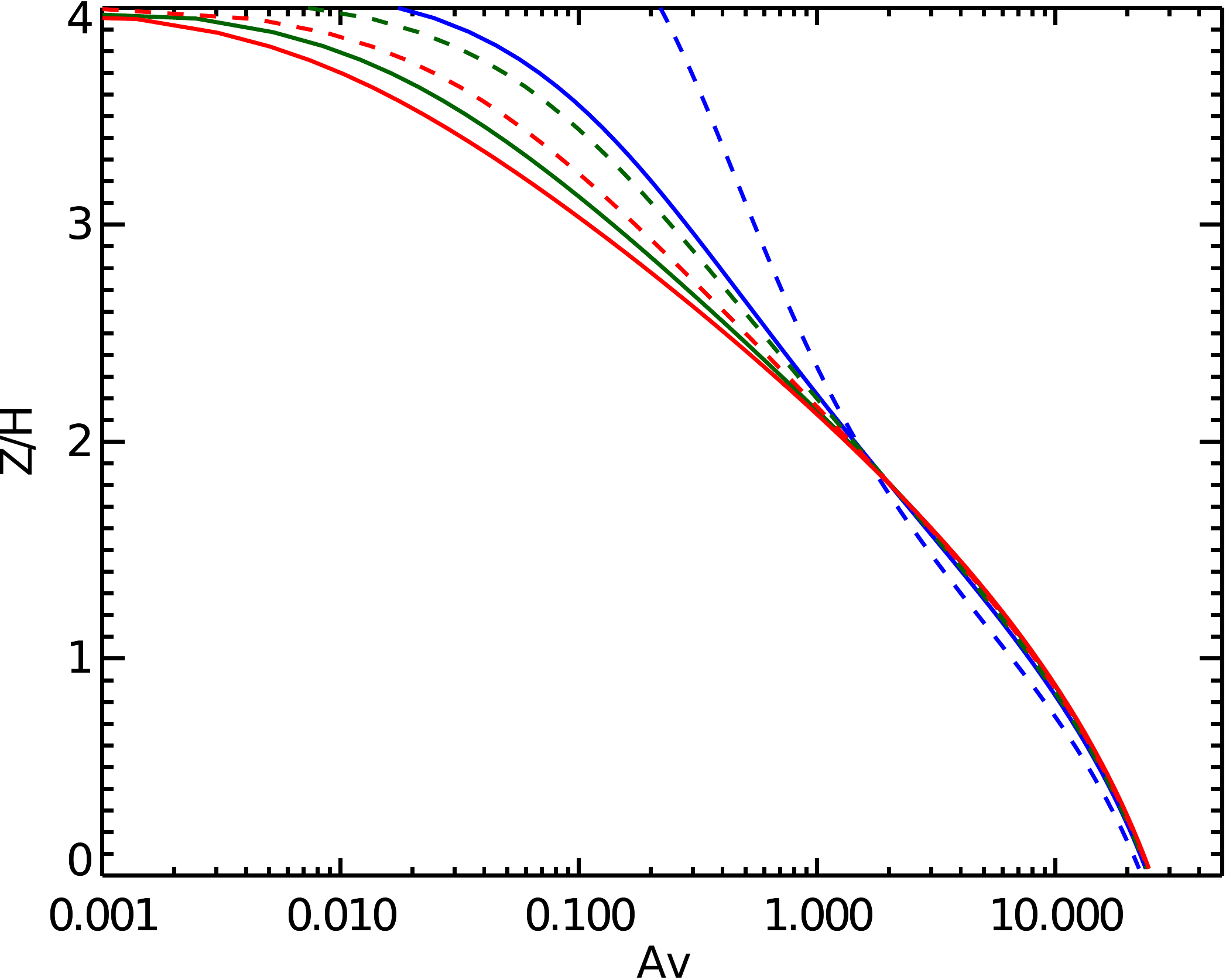}
  \caption{Examples of temperature, density and visual extinction
vertical profiles in the protoplanetary disk at 300 au used for the
chemical modeling. Solid lines are for Model 1 and dashed lines
for Model 2 (see Table \ref{model}). The vertical profiles for
these two models are shown for three different values of the midplane
temperature $T_{\text{mid}}$ at 300 au: $\sim$ 4 K (blue lines),
$\sim$ 20 K (green lines) and $\sim$ 40 K (red lines). Note that the disk heights depend on the temperature: from the coldest to the warmest model, the disk heights are respectively 29, 62 and 88 au at 300 au.}
\label{pardisk}
\end{figure}

For the disk physical parameters, we used the disk model described in \citet{Hersant2009}.

The vertical density structure is obtained by integrating the equation
of hydrostatic equilibrium:
\begin{equation}
\frac{\partial \ln \rho}{\partial z}=-\left[\left(\frac{GM_{*}z}{r^{3}}\right)\left(\frac{\mu m_{\text{H}}}{k_{\text{B}}T}\right)+\frac{\partial \ln T}{\partial z}\right]
\end{equation}
where $G$ is the gravitationnal constant, $M_{*} = 0.5 M_{\odot}$ is
the mass of the central star, $\mu=2.4$ is the reduced mass of total
hydrogen, $m_{\text{H}}$ the mass of the proton and $k_{\text{B}}$
the Boltzmann constant.

For the isothermal case, the vertical density is given by:
\begin{equation}
\rho(z)=\rho_{0} \exp-\left(\frac{1}{2}\left(\frac{z}{H}\right)^{2}\right)
\end{equation}
where $H=(k_{B}T_{\rm mid}r^{3}/GM_{*}\mu m_{\rm H})^{1/2}$ is the
pressure scale height derived from the midplane temperature $T_{\text{mid}}$.

We assumed the disk surface density to vary as $r^{-1.5}$ as described
in \citet{Hersant2009}:
\begin{equation}
\Sigma(r)=\Sigma_{100}\left(\frac{r}{100\,\text{au}}\right)^{-1.5}\,\,\,\, \rm g\, \rm cm^{-2}
\end{equation}
with $\Sigma_{100}=0.8\, \rm g\, \rm cm^{-2}$.
The hydrogen surface density is given by $\Sigma_{100}=\mu m_{\rm H}N_{100}(H)$
in which $\mu$ is the mean weight per H nuclei, which we approximate as $\mu$/2.
Thus $N_{100}(H) \approx 4 \times 10^{23}\,$cm$^{-2}$. With an outer radius of
700 au, the disk mass is $M_{\text{d}}=0.03M_{\odot}$.

Contrary to \citet{Hersant2009}, we use the disk temperature structure
as described in \citet{Williams&Best2014}. The radial midplane
temperature $T_{\text{mid}}$ and the atmospheric temperature
$T_{\text{atm}}$ (temperature at $z \ge 4H$) are given as a power law of the radius:
\begin{equation}
T_{\text{mid}}(r)=T_{\text{mid,1}}\left(\frac{r}{1\,\text{au}}\right)^{-q}
\end{equation}
\begin{equation}
T_{\text{atm}}(r)=T_{\text{atm,1}}\left(\frac{r}{1\,\text{au}}\right)^{-q}
\end{equation}
with $q$\,=\,0.55, which leads to a flaring exponent of 1.23.

A sine function is used for the connection between the midplane and
the atmosphere gas temperatures:
\begin{equation}
T(r,z)=\left\{\begin{array}{l}
T_{\text{mid}}+(T_{\text{atm}}-T_{\text{mid}})\left[\text{sin}\left(\frac{\pi z}{2z_{q}}\right)\right]^{2\delta}\,\,\,\,\text{if}\,\,z < z_{q}\\
T_{\text{atm}} \hspace{4,2cm}\text{if}\,\,z \geq z_{q}
\end{array}
\right.
\end{equation}
with $\delta$\,=\,2 (steepness of the profile) and $z_{q}$\,=\,4$H$
(height at which the disk reaches the atmospheric value).  The visual extinction
is related to the total column density of hydrogen by
$A_{V}=N(H)/1.6\times10^{21}\, \text{cm}^{-2}$, following
\citet{Wagenblast&Hartquist1989}, consistent with the adopted
mean grain radius of 0.1 $\mu$m. Changing the temperature profile also
affects the density and visual extinction distributions.

For the UV radiative transfer in the disk, we consider only the
vertical extinction. We assume that the radial extinction of stellar UV
field is very efficient so that UV photons are scattered by small grains
at high altitudes. The UV flux is decreasing as 1/$r^{2}$ from the central
star. We consider $G_{0}=410$  for the field for a distance of
100 au from the star: this is representative of low mass T Tauri stars like
DM Tau. Half of this flux is assumed to be scattered downwards
inside the disk. To model the CO and H$_2$ UV self-shielding, we used
the approximation from \citet{Lee1996}. We use 64 points in the vertical
direction.

We created a large grid of models by varying the values of
$T_{\text{mid,1}}$ and $T_{\text{atm,1}}$. The disk midplane temperature
is therefore ranging between 4 and 52 K while the temperature in the
upper layer is ranging between 17 and 95 K (at 300 au).
We also considered different
models (see Table \ref{model} for a description of our models). Model 1
is considered as our nominal model. In Model 2 we study the impact of
the steepness of the temperature profile taken as stronger in this
model by increasing the value of $T_{\rm atm,1}$. A third model (Model
3) was considered in which the dust temperature has been considered as
different from the gas one. In that case, the gas temperature is assumed
to follow the strong gradient temperature of Model 2 while the dust
temperature follows the one of Model 1. Fig.\ref{pardisk} shows the disk
temperature, visual extinction and density vertical profiles at 300 au
for some of those models. Depending on the temperature profile considered,
disks have different height, the warmer ones are thicker. From the
coldest to the warmest one shown in Fig.\ref{pardisk}, the disk heights
at 300 au are respectively, 29, 62 and 88 au.


\begin{table}
 \centering
\begin{minipage}{7cm}
 \caption{Models description.}
\begin{tabular}{ll}
\hline \hline
\textbf{Model 1}& \large{$T_{\text{gas}}=T_{\text{dust}}$}\\
&Weak temperature gradient defined as:\\
&\large{$T_{\text{atm,1}}=T_{\text{mid,1}}+300$ {K \footnote{\label{a}for $r$ = 1 au}} }\\\hline
\textbf{Model 2}& \large{$T_{\text{gas}}=T_{\text{dust}}$}\\
&Strong temperature gradient defined as:\\
&\large{$T_{\text{atm,1}}=T_{\text{mid,1}} + 1000$ {K \footref{a}}}\\\hline
\textbf{Model 3}& \large{$T_{\text{gas}}>T_{\text{dust}}$}\\
&Dust temperature profile as in Model 1.\\
&Gas temperature profile as in Model 2.\\\hline
\label{model}
 \end{tabular}
\end{minipage}
 \end{table}

\section{Results}
In this section, we present the results obtained with our models.
We first study the impact of the temperature profile on the disk
chemistry and more particularly on the carbon-bearing species abundances.
Secondly, we show the sensitivity to the initial cloud composition.
The initial chemical conditions used for the disk modeling are
listed in Table \ref{cloud}. Hereafter, for any molecule X, we shall
use s-X to refer to molecules trapped on grains, and X for molecule
in the gas-phase.

\subsection{Impact of the vertical temperature profile}

\begin{figure}[h]
\centering
\includegraphics[width=\columnwidth]{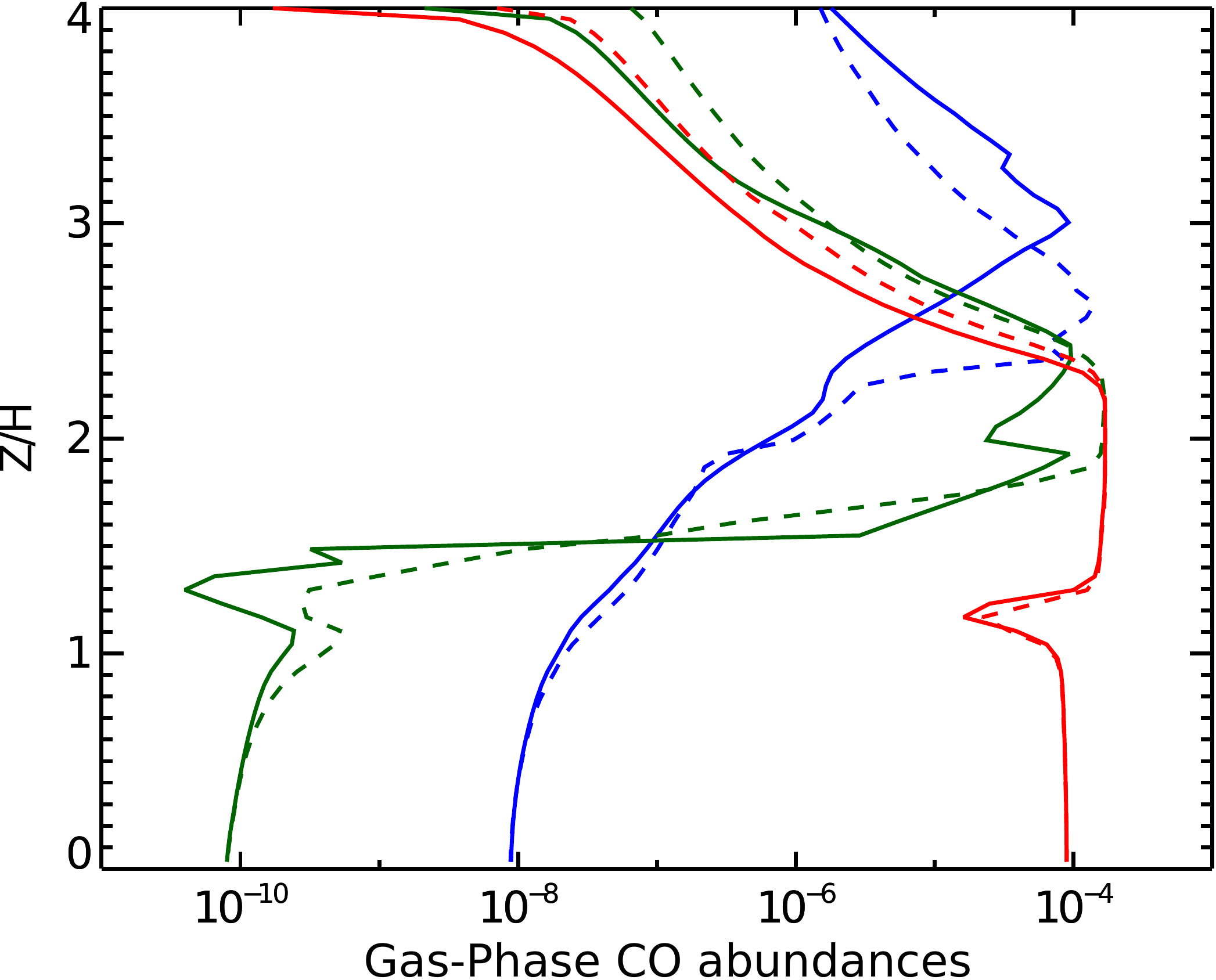}
  \caption{Vertical gas-phase CO abundances (relative to total hydrogen)
  at 300 au for a disk age of 1 Myr.
  The color coding corresponds to the vertical profiles show in Fig.\ref{pardisk}.
  Solid lines are for Model 1 and dashed lines for Model 2 (see Table \ref{model}).}
\label{abz}
\end{figure}

The (gas-phase) CO vertical abundance  at $r = 300$\,au is represented in Fig.\ref{abz} for
the 6 profiles shown in Fig.\ref{pardisk} and for a disk age of 1 Myr. In the upper layer
of the disk, the CO abundance is mostly affected by the photodissociation
process. Because of the different scale heights, the visual extinction at
Z/H=4 is varying between models (see Fig. \ref{pardisk}). In the upper
layer, CO is more efficiently photodissociated by UV photons for the
warmer models (green and red ones) because the visual extinction is smaller.
The largest CO abundances are obtained in the warm molecular layer,
whose height and thickness differs according to the model. The coldest
model with a stronger temperature gradient (blue dashed line in
Fig.\ref{abz}) shows a larger CO abundance at smaller scale height
compared to the coldest model with a weak temperature gradient
(blue solid line in Fig.\ref{abz}). This is due to the fact that
the desorption temperature of CO is reached at smaller height.
In the disk midplane, the larger CO abundance is obtained with the
higher T$_{\text{mid}}$ model (in red in Fig.\ref{pardisk}) because of
a more efficient desorption of s-CO from the ices. For this model,
the CO depletion is small, as expected at such high temperatures. The CO
abundance obtained with the smaller T$_{\text{mid}}$ model (in blue in
Fig.\ref{pardisk})
is higher than the intermediate model (green one). In fact, around 15~K, s-CO is destroyed on the dust grains to form more complex species by secondary
UV photons induced by cosmic-rays, such
as s-CO$_{2}$, s-H$_2$CO and s-CH$_3$OH, whereas when the temperature
is too low (below 9 K), the diffusion of the species, even atomic hydrogen,
at the surface of the grains is less efficient.

The CO abundance profiles obtained with the models are then sensitive to
the vertical temperature profile that is assumed: the midplane temperature
but also the steepness of the gradient. For midplane temperatures larger
than 20~K, the steepness of the gradient no longer substantially
affects the CO abundance because the evaporation temperature of s-CO is
reached in the molecular layer.

For the models displayed in Fig. \ref{abz}, only the warmest ones (red lines)
reach the value CO/H$_2$ = 10$^{-4}$. For the other models, which are
colder, this ratio is much smaller for two main reasons. Firstly, the
warm molecular layer in which larger CO abundances are obtained is
less extensive than in the case of warmer models and secondly, the
CO depletion due to freeze-out and grain surface chemistry is
very efficient in the disk midplane.\\

\begin{figure}[h]
\centering
 \includegraphics[width=\columnwidth]{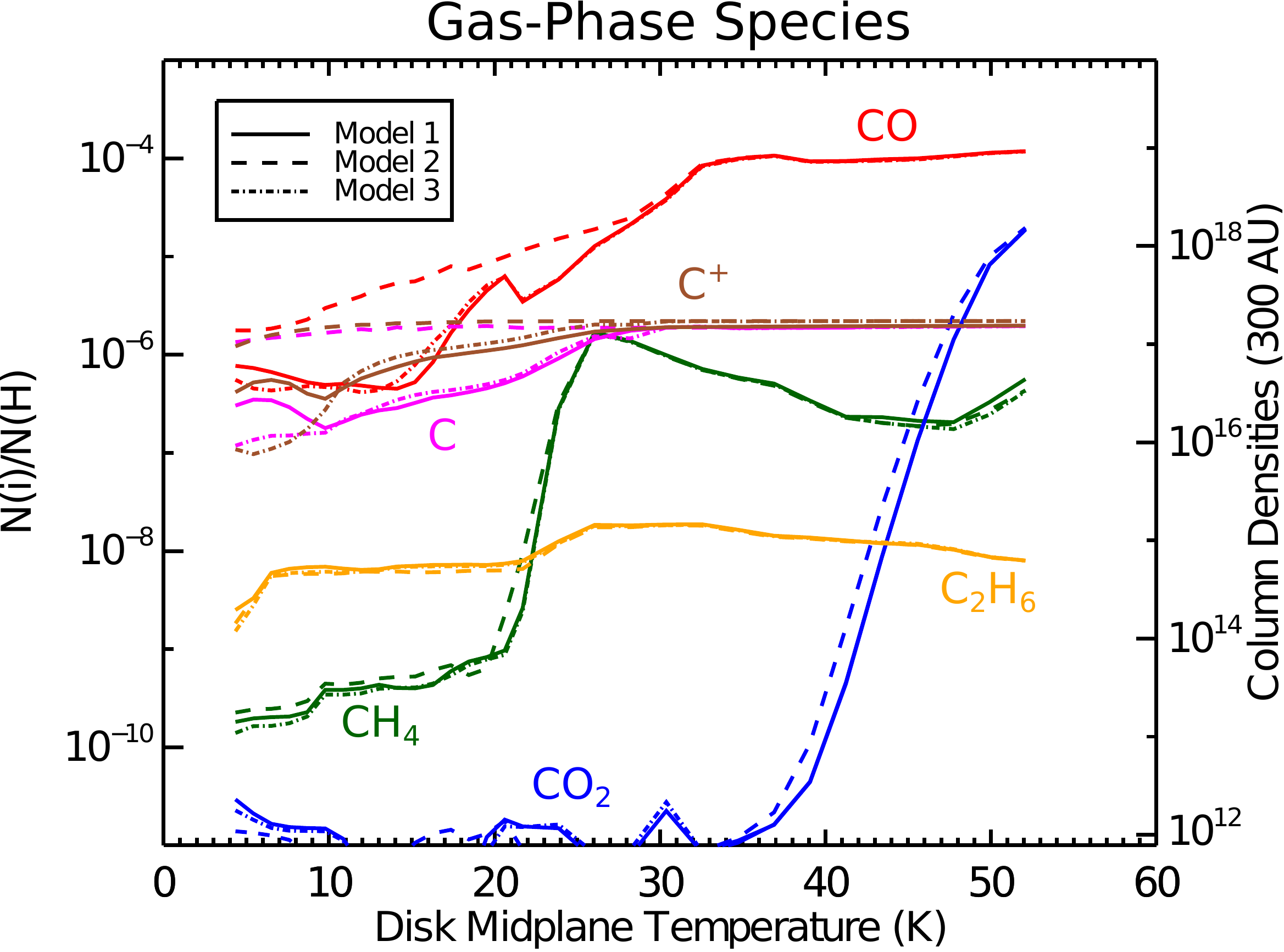}
\vskip2em
    \includegraphics[width=\columnwidth]{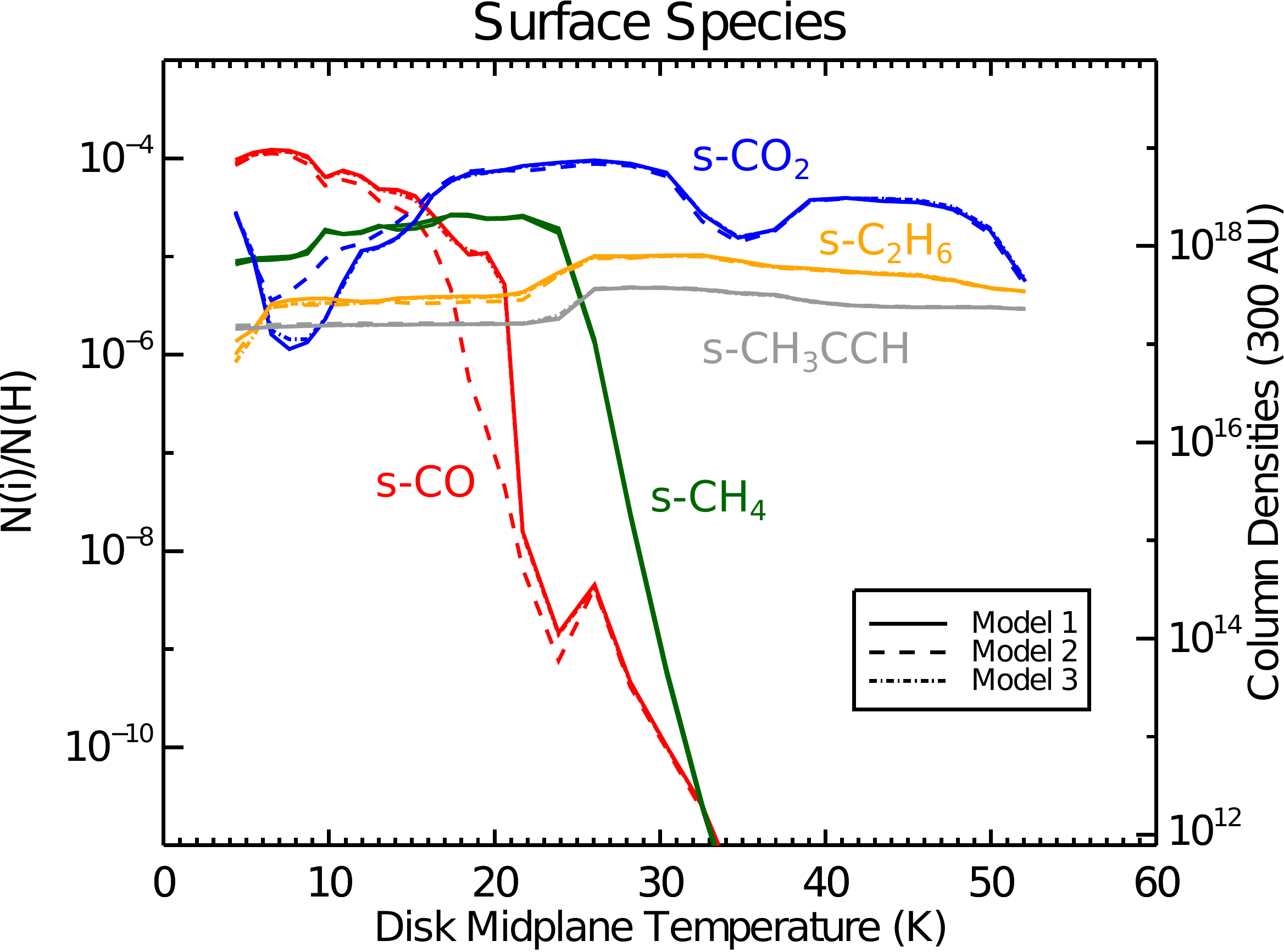}
  \caption{Average abundances and integrated column densities (cm$^{-2}$) of
  carbon-bearing species in the gas-phase and on the grain surfaces for
  Model 1, 2 and 3 as a function of the disk midplane temperature
  (T$_{\text{mid}}$). s- refers for surface species. Those results are
  obtained at 300 au for our nominal cloud and for a disk age of 1 Myr.}
\label{all}
\end{figure}

Figure \ref{all} represents the disk column densities and vertical
abundances (computed from the molecular column densities and the total
H column density) of some carbon-bearing species in the gas-phase and
at the surface of the grains as a function of the disk midplane
temperature and for the three models as described in Table \ref{model}. The results are obtained for a disk age of 1 Myr.

In the case of temperature profiles with a midplane temperature between
5 and $\sim$ 15 K in Model 1, s-CO is the most abundant C-bearing
species. At higher temperature, its abundance decreases because
it is converted into other C-bearing molecules such as s-CH$_4$.
The direct photodissociation of CO in the gas and at the surface of the grains
increases the abundance of neutral carbon, and subsequent hydrogenation
of Carbon on the surfaces contributes to form s-CH$_4$. Above 15 K, s-CO$_{2}$
becomes the most abundant molecule. Its formation
is due to the following reactions, for which the efficiencies are increased
by 10 to $\sim$ 20 orders of magnitude at such temperatures compared to 15 K.
\begin{equation}\label{react1}
\text{s-O} + \text{s-CO}\rightarrow \text{s-CO}_{2}
\end{equation}
\begin{equation}\label{react2}
\text{s-O}+\text{s-HCO}\rightarrow \text{s-CO}_{2}+\text{s-H}
\end{equation}
\begin{equation}\label{react3}
\text{s-OH}+\text{s-CO}\rightarrow\text{s-CO}_{2}+\text{s-H}
\end{equation}
Reaction~\ref{react1} is the main formation path of s-CO$_{2}$ in our model
although it requires to overcome a large activation energy barrier (1000 K).
Reaction~\ref{react2} being barrierless contributes also to form s-CO$_{2}$
efficiently. Reaction~\ref{react3} requires to overcome a small activation
energy barrier (80 K). The warm dust temperature allows to increase the
mobility of species at the surface of the grains and the diffusion becomes
much more efficient. However, the dust temperature is not warm enough to
desorb the s-CO$_{2}$ back into the gas-phase. At even higher temperatures
(above 30~K), CO in gas-phase is the reservoir of C because of
efficient desorption and the lack of sticking on the grains.

Model 3 (in which $T_{\text{gas}} > T_{\text{dust}}$) produces species
abundances very similar to Model 1 whatever T$_{\text{mid}}$, meaning that
it is the grain temperature that is the important parameter here. For
temperature between 25 and 50~K (depending on the molecule), the
steepness of the temperature gradient does not influence the computed
abundances in the gas-phase or at the surface of the grains. For T$_{\text{mid}}$
smaller than 25~K, the gas-phase
abundances are larger for Model 2 mainly because s-CO desorbs from
the grains in Model 2 in the molecular layer, producing the other
C-bearing species, while it stays on the grains in Model 1.

s-CH$_3$OH and s-H$_2$CO are efficiently formed at low
T$_{\rm mid}$: between 10 and 20 K, their abundances are respectively,
$\sim$ 10$^{-5}$ and $\sim$ 10$^{-6}$. Their main formation paths are:
\begin{equation}
\text{s-OH} + \text{s-CH}_{3}\rightarrow \text{s-CH}_{3}\text{OH}
\end{equation}
\begin{equation}
\text{s-H} + \text{s-CH}_{2}\text{OH}\rightarrow \text{s-CH}_{3}\text{OH}
\end{equation}
\begin{equation}
\text{s-H}+\text{s-HCO}\rightarrow\text{s-H}_{2}\text{CO}
\end{equation}
\begin{equation}
\text{s-O}+\text{s-CH}_{2}\rightarrow\text{s-H}_{2}\text{CO}
\end{equation}
However, their abundances are strongly decreased at higher T$_{\rm mid}$
since s-H, s-HCO, s-O and s-CH$_{2}$ are easily desorbed at such
temperatures. The largest CH$_3$OH and H$_2$CO gas-phase column densities are
found to be around 10$^{12}$ cm$^{-2}$ for T$_{\rm mid}$ = 20 K.\\

In summary, the carbon reservoir in protoplanetary disks strongly depends on the
disk vertical temperature profile which is controlled by the midplane temperature;
the reservoirs are:
\begin{itemize}\itemsep 0pt
\item[-] s-CO in the ices below 15~K.
\item[-] s-CO$_2$ in the ices between 15 and 30~K.
\item[-] CO in the gas-phase above 30~K.
\end{itemize}

\begin{figure}[h]
\centering
 \includegraphics[width=\columnwidth]{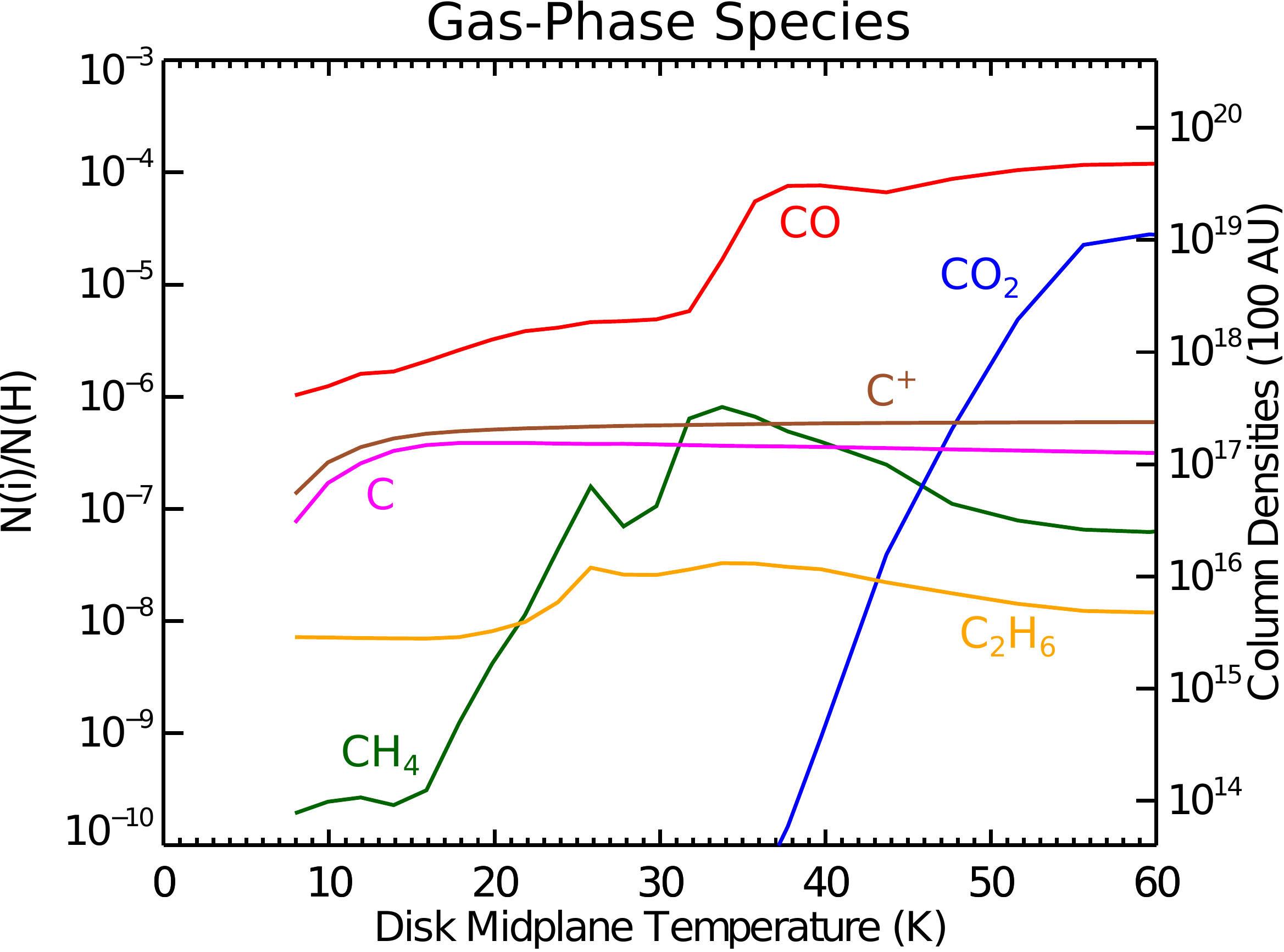}
\vskip2em 
    \includegraphics[width=\columnwidth]{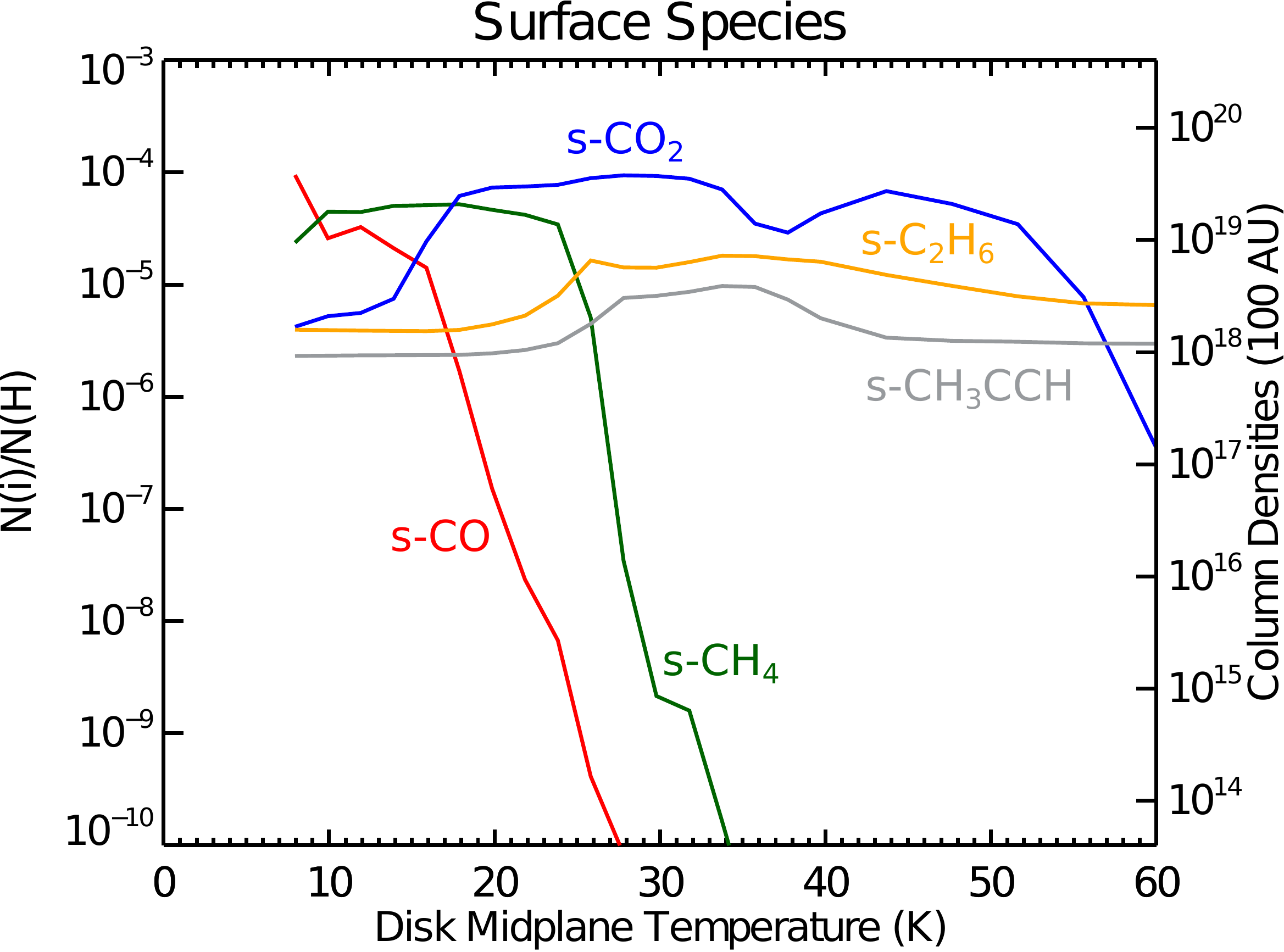}
  \caption{Average abundances and integrated column densities (cm$^{-2}$) of
  carbon-bearing species in the gas-phase and on the grain surfaces at
  100 au as a function of the disk midplane temperature (T$_{\text{mid}}$). Results are
  for Model 1 and for a disk age of 1 Myr.}
\label{100au}
\end{figure}

\begin{table}
\centering
  \caption{Molecular binding (desorption) energies.}
  \begin{tabular}{lc}
  \hline \hline
Species&$E_{D}$ (K)\\
\hline
C& 800\\
CO & 1150\\
CH$_{4}$&1300\\
H$_2$CO&2050\\
CO$_{2}$& 2575\\
CH$_{3}$CCH& 3837\\
C$_{2}$H$_{6}$&4387\\
CH$_3$OH&5534\\
\hline
\label{ED}
\end{tabular}
\end{table}

The results described above were obtained at 300 au. We show now
the results of Model 1 (weak temperature gradient) at 100 au. In Fig.\ref{100au},
are represented
the disk abundances and column densities of carbon-bearing species in
the gas-phase and on the grain surfaces at 100 au for Model 1 as a
function of the disk midplane temperature (to be compared to Fig.\ref{all}).
At 100 au, for a given vertical temperature profile, the density,
the visual extinction, and the UV fluxes are larger than at 300 au.
Considering the larger densities, the reservoirs of carbon at low
temperature (T$_{\rm mid} \le 17$~K) are s-CH$_4$ and s-CO, and
s-CO$_2$ between 17 and 34~K. At larger
temperatures, gas-phase CO becomes dominant.

\subsection{Photodesorption}

Photodesorption was not included in our previous models. High desorption yields were found by \citet{Oberg2007}, but it has been shown that the desorption efficiency depends on the shape of the UV spectrum \citep{Fayolle2011}. In particular, the photodesorption yield of CO ice induced by Ly-$\alpha$ photons is small \citep{Chen2014}.

We tested the impact of photodesorption on our results at 100 and 300 au for a disk age of 1 Myr. We used a large yield of 10$^{-3}$ molecules photons$^{-1}$ for all species except for CH$_3$OH, H$_2$O, CO$_2$, N$_2$, and CO for which we used the formalism described in \citet{Oberg2009a, Oberg2009b, Oberg2009c}. 

The carbon reservoir remains the same for both radii. The gas-phase CO abundance is only affected at 300 au for midplane temperatures between 4 and 15 K, with a CO abundance five times larger than in our nominal model. However, s-CO remains the main carbon-bearing species. For both radii, only more complex molecules, such as C$_2$H$_6$ and CH$_3$CCH, are efficiently photodesorbed (their abundances are increased by more than a factor 10).

\subsection{Impact of the cosmic ray ionization rate}
Cosmic rays are important for the ionization in the disk. However, as recently
suggested by \citet{Cleeves2013}, cosmic rays penetration may be highly reduced in
the presence of stellar winds and/or magnetic fields. In this section
we study the impact of a lower (cosmic ray) ionization rate on the disk chemistry.

\begin{figure}[h]
\centering
 \includegraphics[width=\columnwidth]{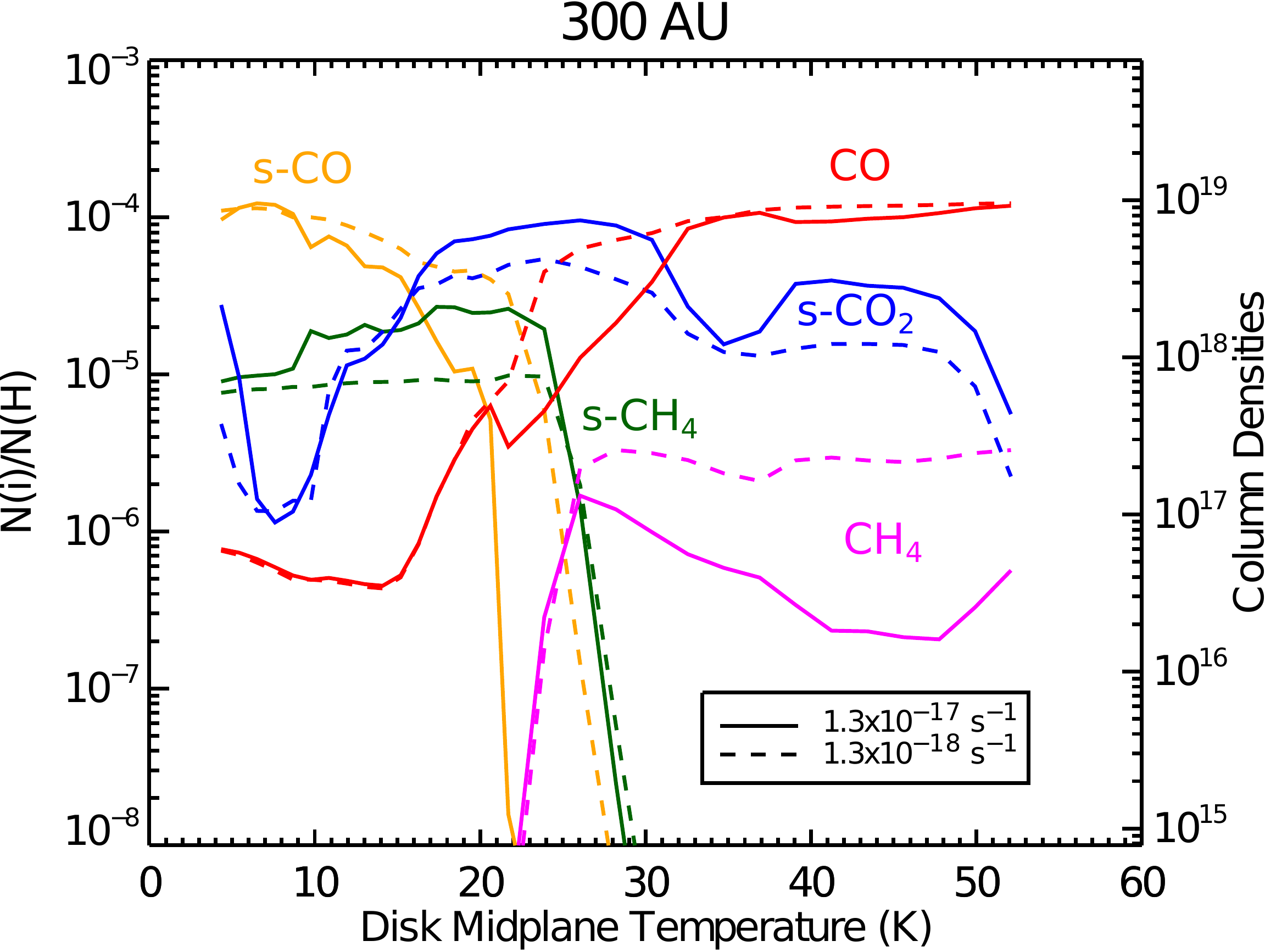}
\vskip2em 
    \includegraphics[width=\columnwidth]{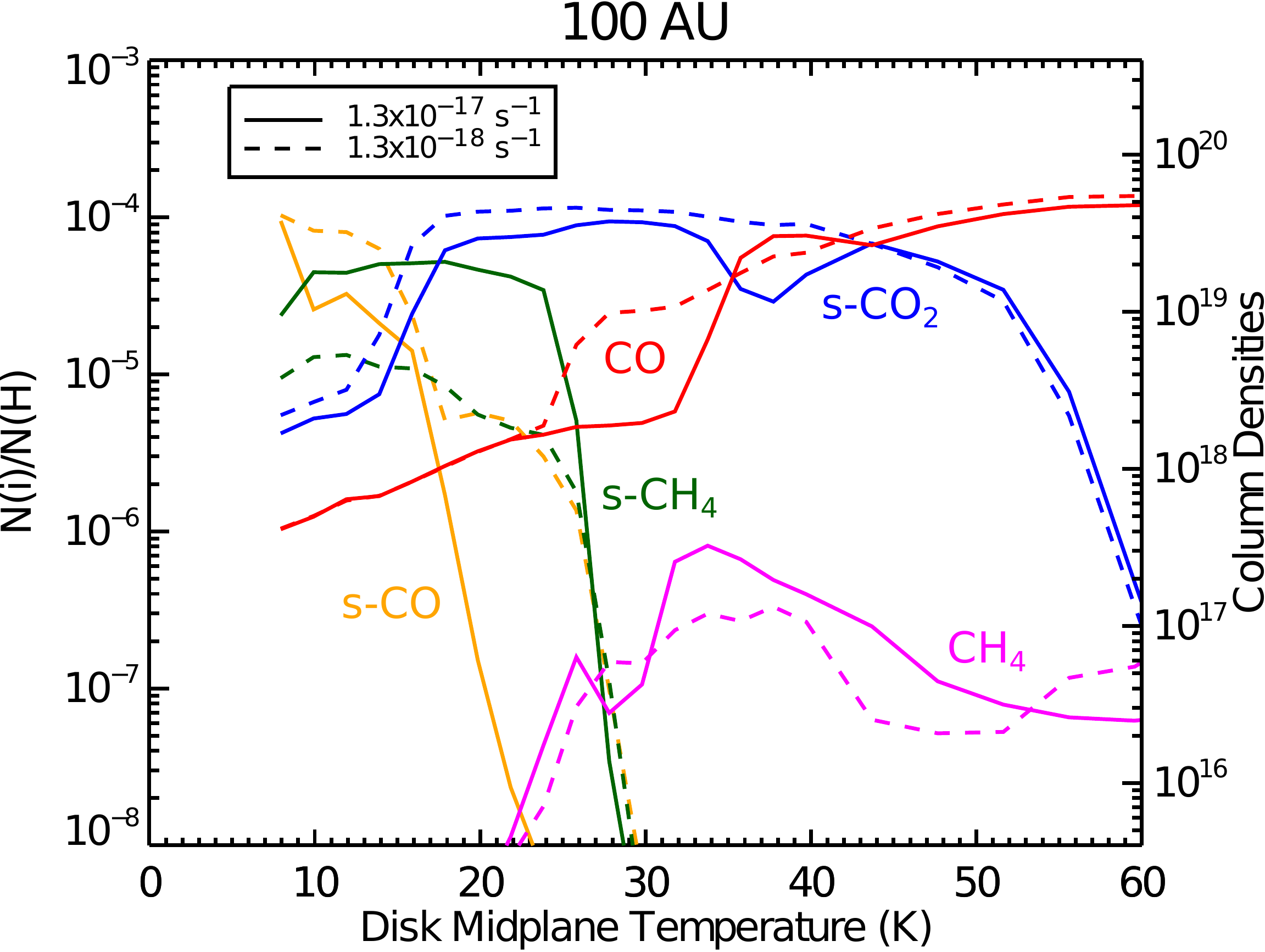}
  \caption{Average abundances and integrated column densities (cm$^{-2}$) of
  main reservoirs of carbon at 300 au and 100 au for two values of cosmic-rays
  ionization rate as a function of the disk midplane temperature
  (T$_{\text{mid}}$). Results are for Model 1 and for a disk age of 1 Myr.}
\label{crir}
\end{figure}

Figure \ref{crir} shows the abundances of the main Carbon-bearing species
for two values of
the cosmic rays ionization rate: 1.3$\times$10$^{-17}$ and
1.3$\times$10$^{-18}$ s$^{-1}$ (respectively in solid and dashed lines).
The impact of ionization rate on CO abundance is limited, being restricted
to the temperature range 20 -- 35 K. Considering the lower value for $\zeta_{\text{CR}}$, s-CO abundance increases because its dissociation on the grain surface by UV photons induced by cosmic rays is less efficient. As a consequence, there is less carbon available to form s-CH$_4$. In the gas-phase, the dissociation by secondary UV photons induced by cosmic ray/H$_2$ interactions is also less efficient so that CH$_4$ abundance increases at 300 au. At 100 au, the UV flux being larger, the direct photodissociation is the dominant process. CH$_4$ presents therefore similar sensitivity than s-CH$_4$ and decreases as well.

We also ran our Model 1 for $\zeta_{\text{CR}}$ =
1.3$\times$10$^{-19}$ s$^{-1}$ but the reservoirs remain the
same as those shown in Fig.\ref{crir} (dashed lines) for both disk radii.

\subsection{Impact of the age of the disk}

The conversion of s-CO into more complex molecules on grains is a
time dependent process, which requires a minimum timescale to occur.
Up to now, the results described were obtained for disks of 1 Myr. In
this section we study the chemistry of carbon-bearing species for disks
5 times older.

\begin{figure}[h]
\centering
 \includegraphics[width=\columnwidth]{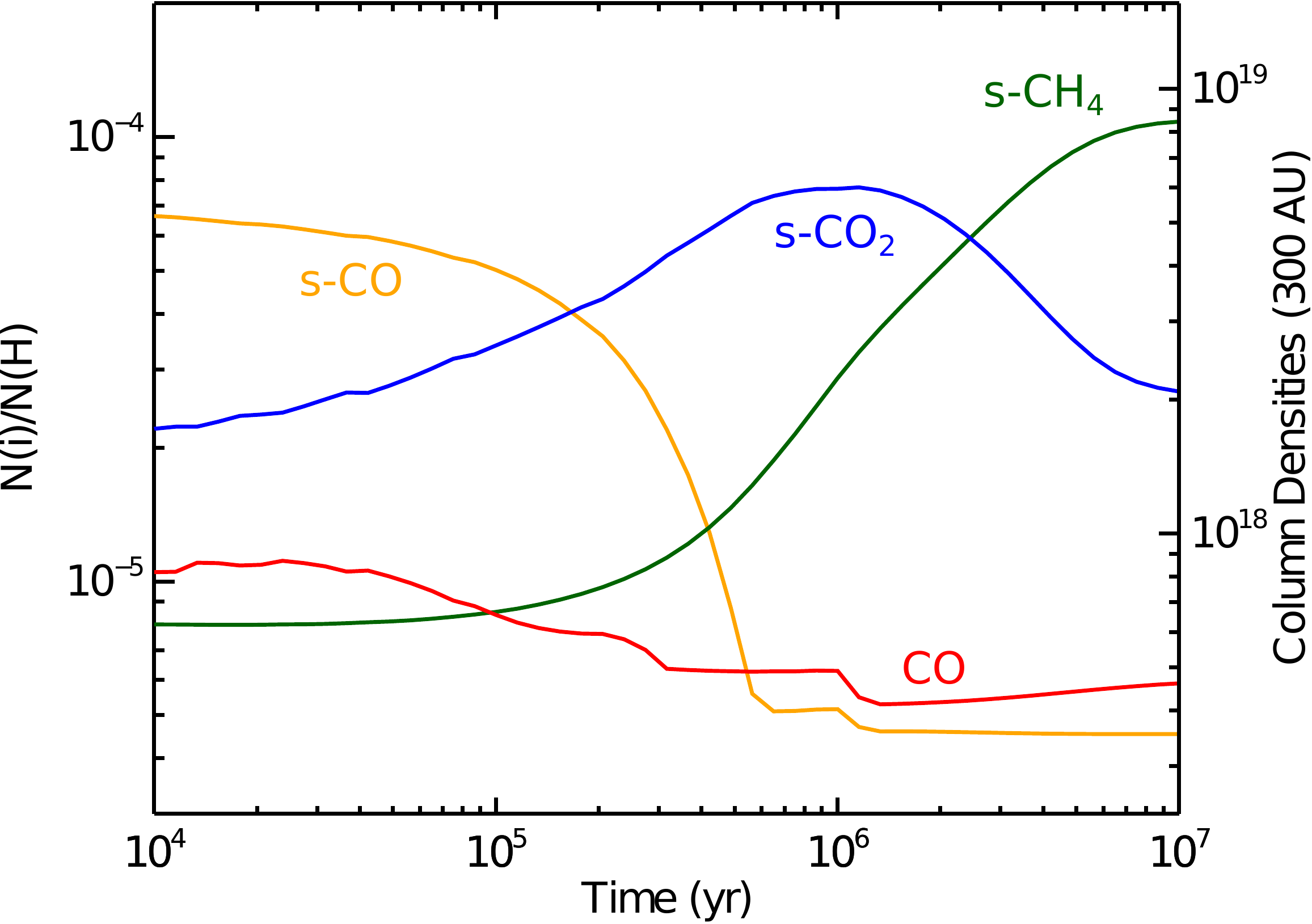}
  \caption{Average abundances and integrated column densities (cm$^{-2}$)
  of main reservoirs of carbon at 300 au as a function of time. The
  results are obtained for Model 1 and for $T_{\text{mid}}$ = 20 K at 300 au.}
\label{abvsage}
\end{figure}

Figure \ref{abvsage} shows the abundances of CO, s-CO, s-CO$_2$, and s-CH$_4$, obtained for Model 1 and for $T_{\text{mid}}$ = 20 K, as a function of time. The s-CO abundance strongly decreases at 10$^{5}$ yr due to its conversion into other species such as s-CO$_2$ and  s-CH$_4$. We can see that the formation of s-CH$_4$ requires a longer timescale to occur than the formation of s-CO$_2$. s-CH$_4$ reaches its highest abundance at 10$^{7}$ yr while s-CO$_2$ is the most abundant species for a disk at age of 10$^{6}$ yr. The reservoirs of carbon are therefore changed while considering an older disk.

\begin{figure}[h]
\centering
 \includegraphics[width=\columnwidth]{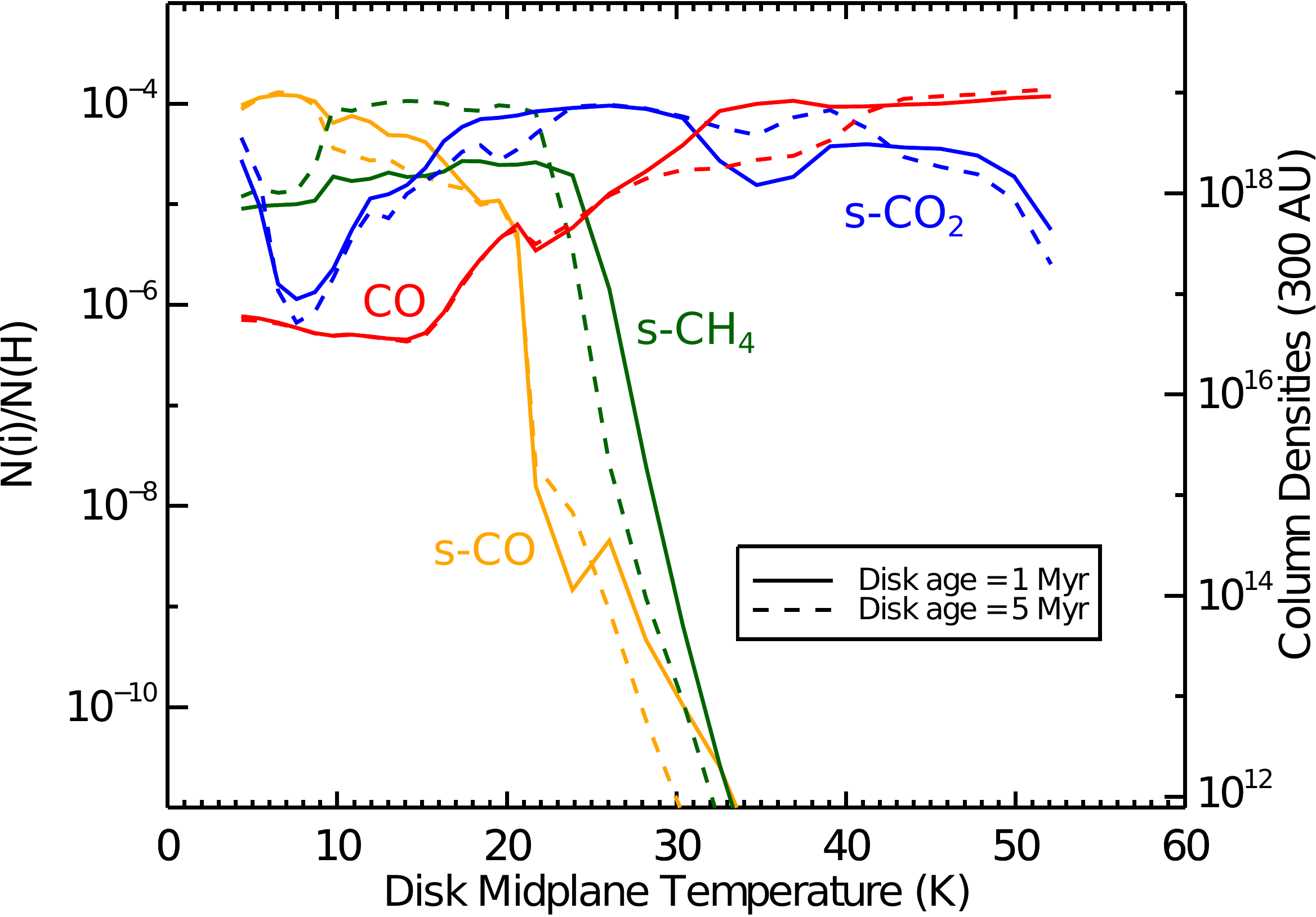}
  \caption{Average abundances and integrated column densities (cm$^{-2}$) of
  main reservoirs of carbon at 300 au for two disk ages as a function of
  the disk midplane temperature (T$_{\text{mid}}$). Results are obtained
  for our Model 1.}
\label{vsage}
\end{figure}

Figure \ref{vsage} represents the abundances and column densities of the main carbon-bearing species as a function of the disk midplane temperature for disks of 1 and 5 Myr. For the older disk, s-CH$_4$ is now the reservoir of carbon between 10 and 22 K in the midplane which is due to its long formation timescale. s-CO$_2$ is also more abundant than CO between 32 and 40 K at 5 Myr compared to 1 Myr. This apparent dependency with temperature
is actually a density effect. Because of the hydrostatic scale height,
densities in the disk are smaller for higher temperatures, and the
conversion time is just inversely proportional to the density. The
conversion of CO into s-CO$_2$ is then longer in warmer disk so that
at 35 K CO is the main carbon-bearing species for a disk of 1 Myr. For
instance, in our model, this conversion timescale varies from
$6 \times 10^5$ yr for disks at 25 K to $2 \times 10^6$ yr for
disks at 35 K.

\subsection{Sensitivity to the initial conditions}
We now study the impact of the chemical composition of the parent molecular
cloud on the disk chemistry. Several parameters have been studied such
as gas density, C/O elemental ratio and age of the parent cloud. The
abundances of the main C-bearing species obtained for different cloud
parameters, and used as initial conditions for the disk chemistry, are
listed in Table \ref{cloud}. The results presented in this section are for disks of 1 Myr.

\subsubsection{Density of the parent molecular cloud}

To test the importance of the density of the parent molecular cloud,
we redid our simulations but multiplied the density of the parent cloud
by a factor of ten. The disk carbon chemistry is not very sensitive to
the gas density of the parent molecular cloud. The reservoirs of carbon
as described in the previous section are not changed. Only CH$_3$CCH
abundance in the ices is decreased whatever T$_{\rm mid}$ by a factor
up to $\sim$ 2.

\subsubsection{C/O elemental ratio}
\label{sec:sub:elemental}
For the carbon abundance, we use the value determined by
\citet{Jenkins2009} and we vary the C/O ratio by considering two different
values for the oxygen elemental abundance: 1) 3.3$\times$10$^{-4}$ (value
of our nominal model), a low depletion case, and 2) 1.4$\times$10$^{-4}$,
a high depletion case (see section 2.1.2).\\

The reservoirs of carbon are not changed while using the high depletion
case for C/O elemental ratio (C/O = 1.2). The only main difference is a
smaller abundance of CO containing species and a larger abundance of CH
containing species under some conditions. At 7~K for instance, the CO,
CH$_4$ and CO$_2$ abundances in the ices are respectively
8.8$\times$10$^{-5}$, 2.5$\times$10$^{-5}$ and 3.8$\times$10$^{-7}$.

\subsubsection{Age of the parent molecular cloud}

\begin{figure}
\centering
      \includegraphics[width=\columnwidth]{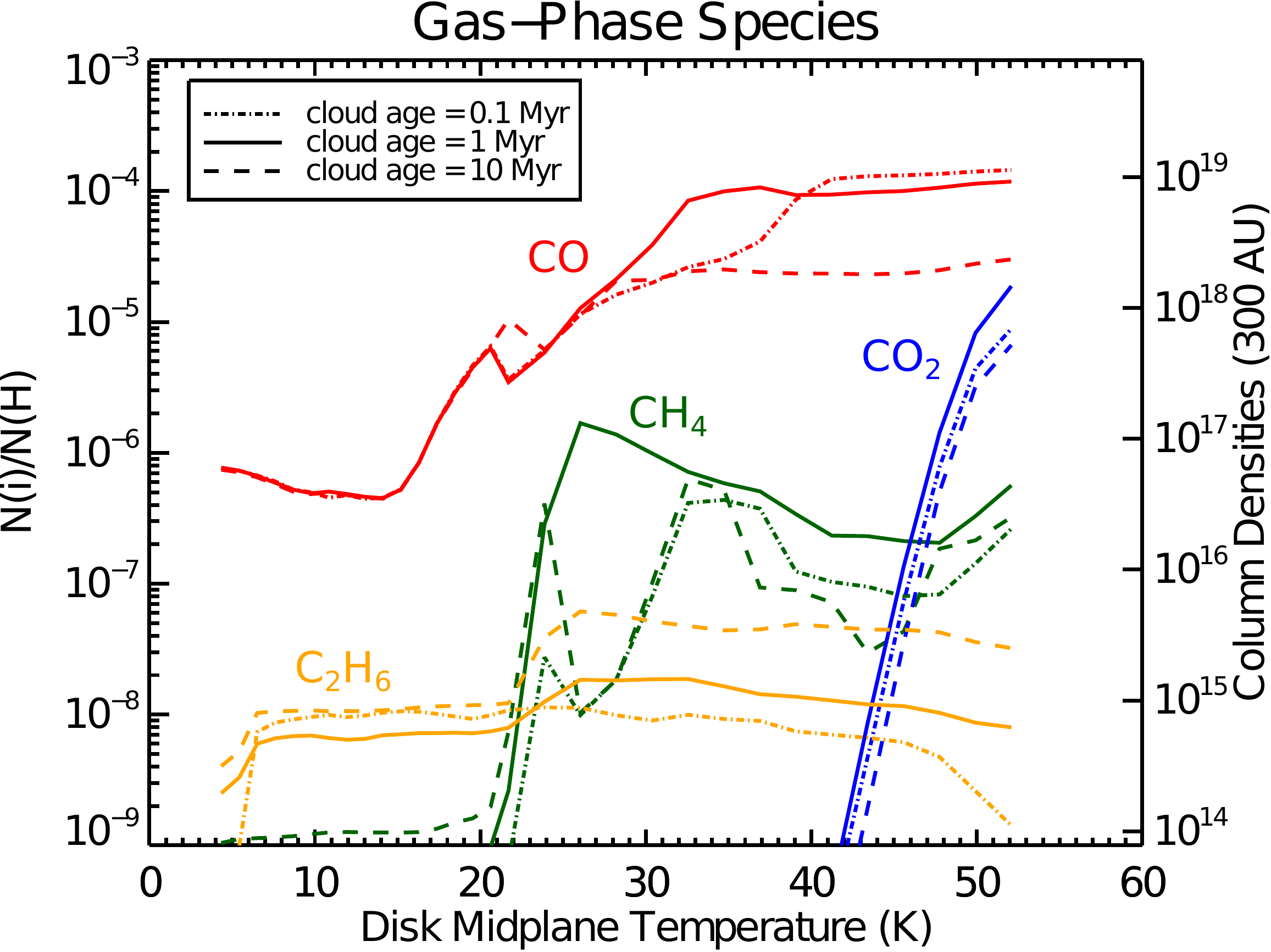}
\vskip2em
      \includegraphics[width=\columnwidth]{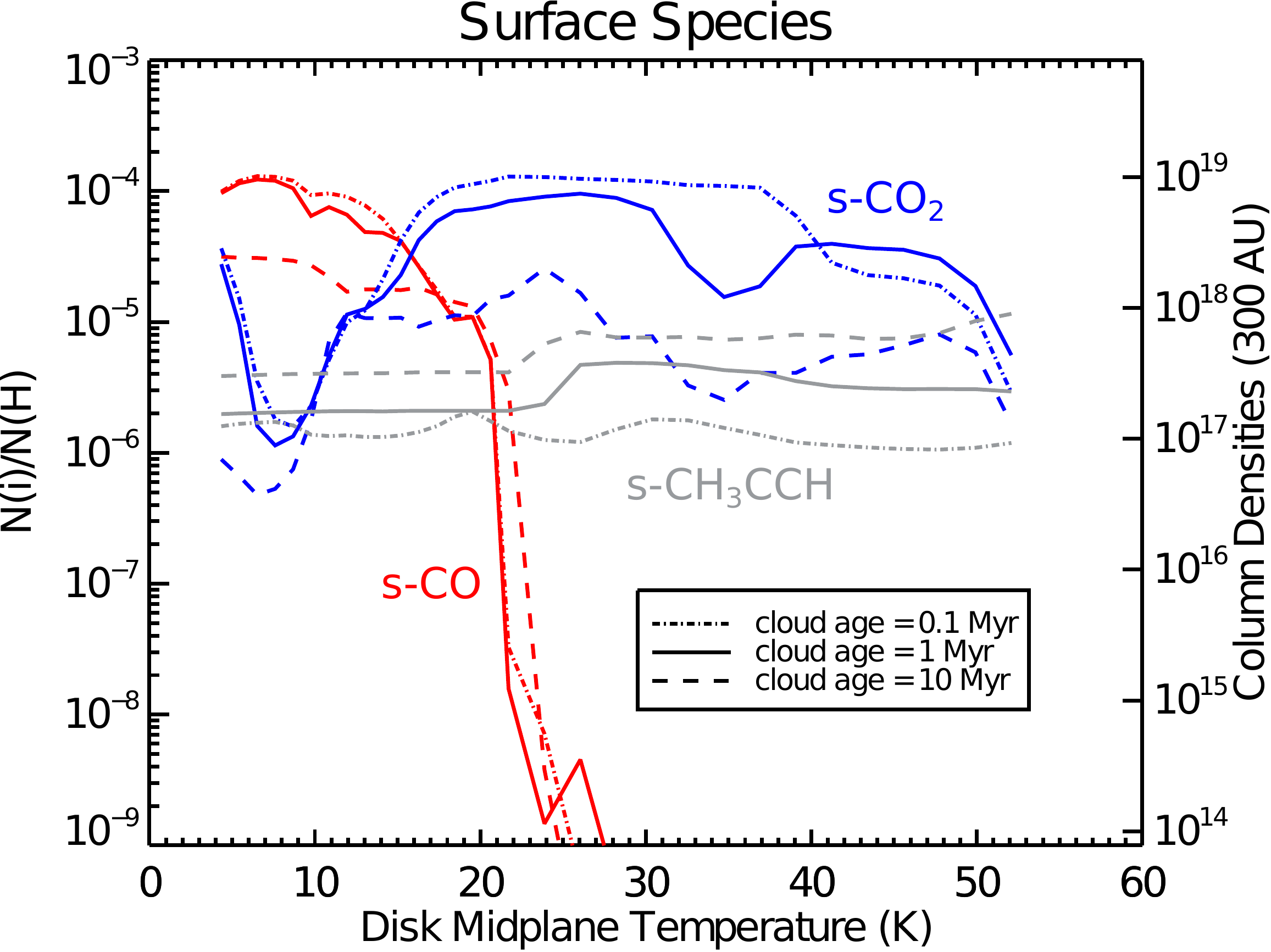}
\vskip2em
      \includegraphics[width=\columnwidth]{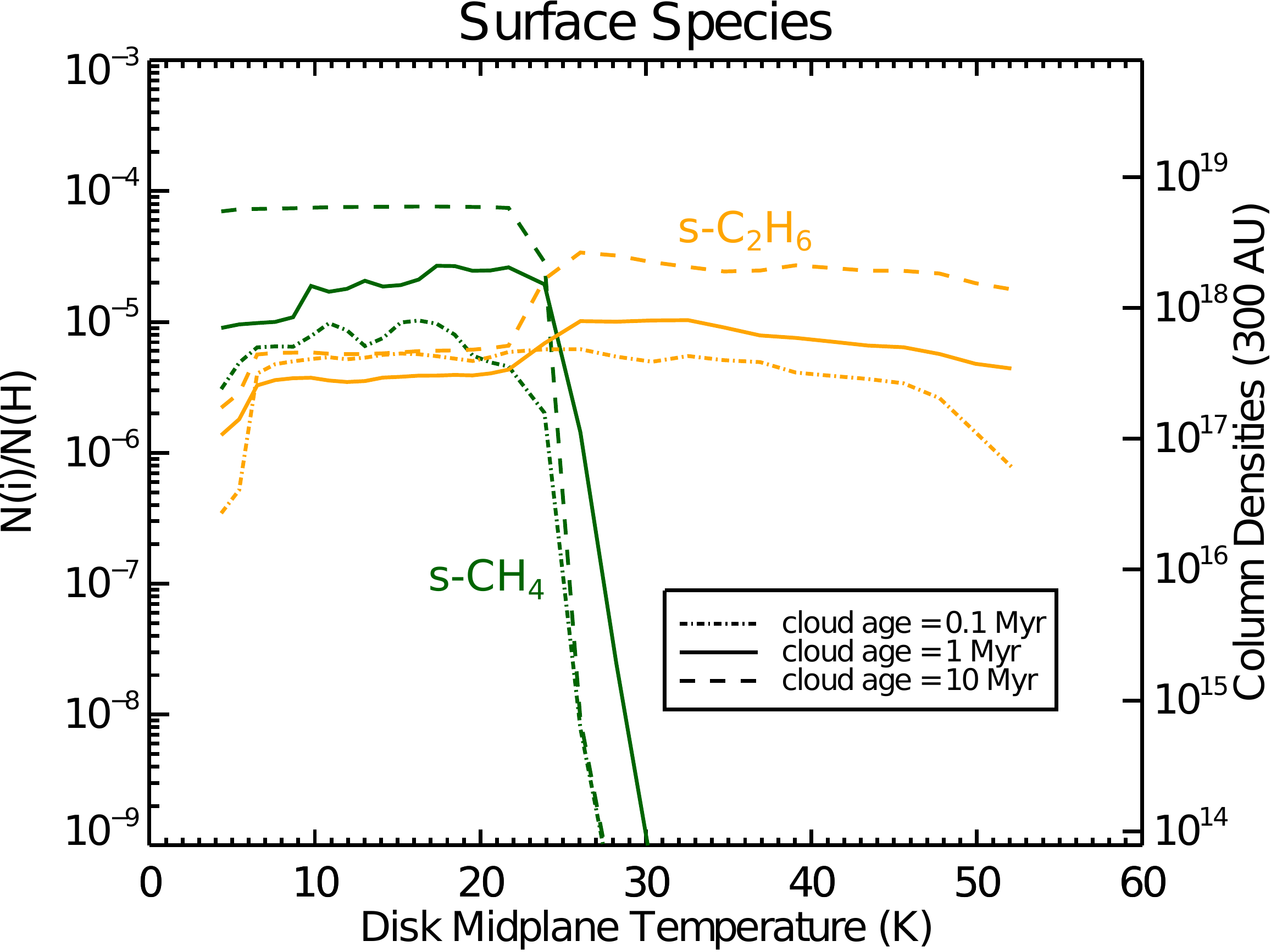}
\caption{Average abundances and integrated column densities (cm$^{-2}$) of
carbon-bearing species in the gas-phase and on the grain surfaces for
three ages of the parent molecular cloud  as a function of the disk
midplane temperature. Results are for Model 1 at 300 au, and for a disk age
of 1 Myr. The results obtained for C and C$^{+}$ are
not displayed since the differences are very small.}
\label{cloudage}
\end{figure}

The age of molecular clouds is still a subject of debate. Ages are
often considered between a few million years
\citep{Hartmann2001,Ballesteros&Hartmann2007} and 10 million years
\citep{Mouschovias2006}. \citet{Ballesteros&Hartmann2007} suggested
a short timescale of a few million years. They argued that starless
molecular clouds being rarely observed, star formation should occur
shortly after cloud formation. An argument favouring long cloud
lifetime is the difficulty to explain the sufficient formation of
molecular hydrogen in only a few Myr considering the slow rate at which
H$_2$ forms in the ISM \citep{Hollenbach1971}. However,
\citet{Glover&MacLow2007} suggested that turbulence significantlly
reduces the time required for the conversion of H into H$_2$
(less than 1-2 Myr). \citet{Pagani2011} suggested that the absence
of DCO$^+$ in dark clouds is due to a high ortho-H$_2$ abundance
which provides an upper limit to the age of 3-6 million years.
More recently, \citet{Brunken2014} used the abundance ratio of
ortho-H$_2$D$^+$ and para-H$_2$D$^+$ and determined an age of dense
molecular clouds around one million year.

To study the sensitivity of the model results to the age of the
parent molecular cloud, we have considered two additional cases.
We computed the initial cloud composition at $10^5$~yr and $10^7$~yr,
in addition to our standard model at $10^6$~yr. In Fig.\ref{cloudage},
we display the abundances and column densities of carbon-bearing species
as a function of the disk midplane temperature computed with the
different initial compositions. The abundances of the main C-bearing
species obtained in the parent cloud at the different times ($10^5$,
$10^6$ and $10^7$~yr) are listed in Table~\ref{cloud}. The reservoirs
of carbon in the disk are the same if a younger parent cloud is
considered. However CO reaches its canonical value at larger
$T_{\text{mid}}$: 40 K instead of 32 K. In the case of an older cloud
the reservoirs of carbon are changed.
At temperatures smaller than 9~K for instance and in the case of a
$10^7$~yr cloud, CO in the ices is not the main C-bearing molecule
anymore while CH$_4$ in the ices is. This can be explained by
differences in the initial chemical composition of the cloud.
The $10^5$ and $10^6$~yr compositions are similar and only the
fraction between gas-phase and surface CO is reversed. At the
large density of the disks, the gas-phase CO is quickly depleted
on the grains. At $10^7$~yr in the cloud, the CO in the ices is
hydrogenated to form H$_2$CO and CH$_3$OH in the ices but is also
photodissociated by secondary UV photons induced by cosmic-rays and the
atomic carbon is hydrogenated to s-CH$_4$.
So that the reservoir of carbon in the disk remains CH$_4$ in the ices.
At larger temperature, s-CO$_2$ is not anymore the reservoir of carbon
since CO$_2$ is formed on the surfaces from CO (already under abundant
in the cloud) and the reservoir is C$_2$H$_6$ in the ices. As a summary, the reservoirs of carbon
in a disk formed from an older molecular cloud with an age of $10^7$~yr
are CH$_4$ in the ices for temperature profiles computed with a
T$_{\text{mid}}$ < 22 K and C$_2$H$_6$ in the ices for larger T$_{\text{mid}}$.\\

The reservoir of carbon in disks is thus rather insensitive to
initial conditions. The age of the cloud has an effect only if it
is old enough, at least 10 Myr, a value on the high end of
estimated cloud ages. In that case, s-CH$_4$ becomes
the dominant C-bearing molecule instead of s-CO in the ices
(for T$_{\text{mid}}$ < 22 K). For instance, for $T_{\text{mid}}$ = 15 K,
the s-CH$_4$/s-CO ratio is $\sim$ 5. Such a ratio is much larger
than those found in Solar System comets \citep[see Table 1 of][]{Bockelee-Morvan+etal_2004},
which may indicate that our Solar System originates from a younger cloud.

\begin{table*}[h!]
 \centering
 \begin{minipage}{12cm}
  \caption{Predicted abundances in the molecular cloud considering different model parameters.}
  \begin{tabular}{lccccc}
  \hline \hline
  &\multicolumn{5} {l} {{Abundances \footnote{a(b) represents a$\times$10$^{b}$}} (with respect to n$_{\text{H}}$)}\\ \hline
species & nominal cloud  & C/O=1.2  & 10$^{5}$ yrs & 10$^{7}$ yrs&n$_{\text{H}}$=2$\times$10$^{5}$\,cm$^{-3}$ \\\hline
CO & 2.4(-5)&1.3(-5)&7.2(-5)&2.9(-7)&2.2(-6)\\
CH$_{4}$& 1.5(-7)&4.0(-7)&4.5(-7)&6.5(-7)&2.4(-8)\\
CO$_{2}$& 3.3(-8)&6.0(-9)&4.7(-8)&9.2(-11)&7.9(-10)\\
C& 3.1(-8)&7.0(-8)&2.9(-5)&5.4(-8)&1.4(-8)\\
C$^{+}$&4.2(-9)&8.0(-10)&2.0(-9)&1.2(-13)&4.4(-10)\\
C$_{2}$H$_{6}$&6.2(-9)&7.1(-9)&7.1(-11)&1.2(-8)&8.3(-9)\\
H$_2$CO&9.8(-9)&1.8(-8)&4.5(-8)&1.6(-9)&2.3(-9)\\
CH$_3$OH&7.1(-10)&1.0(-9)&7.2(-11)&6.2(-11)&1.1(-10)\\
s-CO& 7.2(-5)&3.0(-5)&2.5(-5)&1.9(-7)&9.9(-5)\\
s-CH$_{4}$& 8.5(-6)&2.5(-5)&4.2(-6)&9.2(-5)&5.1(-6)\\
s-CO$_{2}$& 1.6(-6)&5.6(-8)&8.1(-7)&3.9(-9)&2.3(-6)\\
s-C$_{2}$H$_{6}$& 3.4(-6)&3.9(-6)&3.9(-8)&6.4(-6)&4.6(-6)\\
s-H$_2$CO&2.8(-5)&4.1(-5)&1.8(-6)&1.0(-6)&2.3(-5)\\
s-CH$_3$OH&9.7(-6)&1.6(-5)&5.0(-7)&2.5(-6)&5.8(-6)\\
s-CH$_{3}$CCH& 2.4(-6)&6.1(-6)&1.0(-6)&4.6(-6)&1.0(-6)\\
\hline
\label{cloud}
\end{tabular}
\end{minipage}
\end{table*}

\section{Discussion}

Transformation of CO into more complex species on grain surfaces
was already invoked by \citet{Aikawa1997}. However, in their study,
the process is driven by neutral carbon and required enhanced X-ray
ionization to be started. Ionized helium atoms react with CO to release
C, which further react on grains to get CO$_2$ or other hydrocarbons.
We have shown here that with a more complete grain chemistry, this
mechanism is no longer necessary, and CO can be converted to other
forms even without enhanced ionization.

\begin{figure}[h]
\centering
\includegraphics[width=\columnwidth]{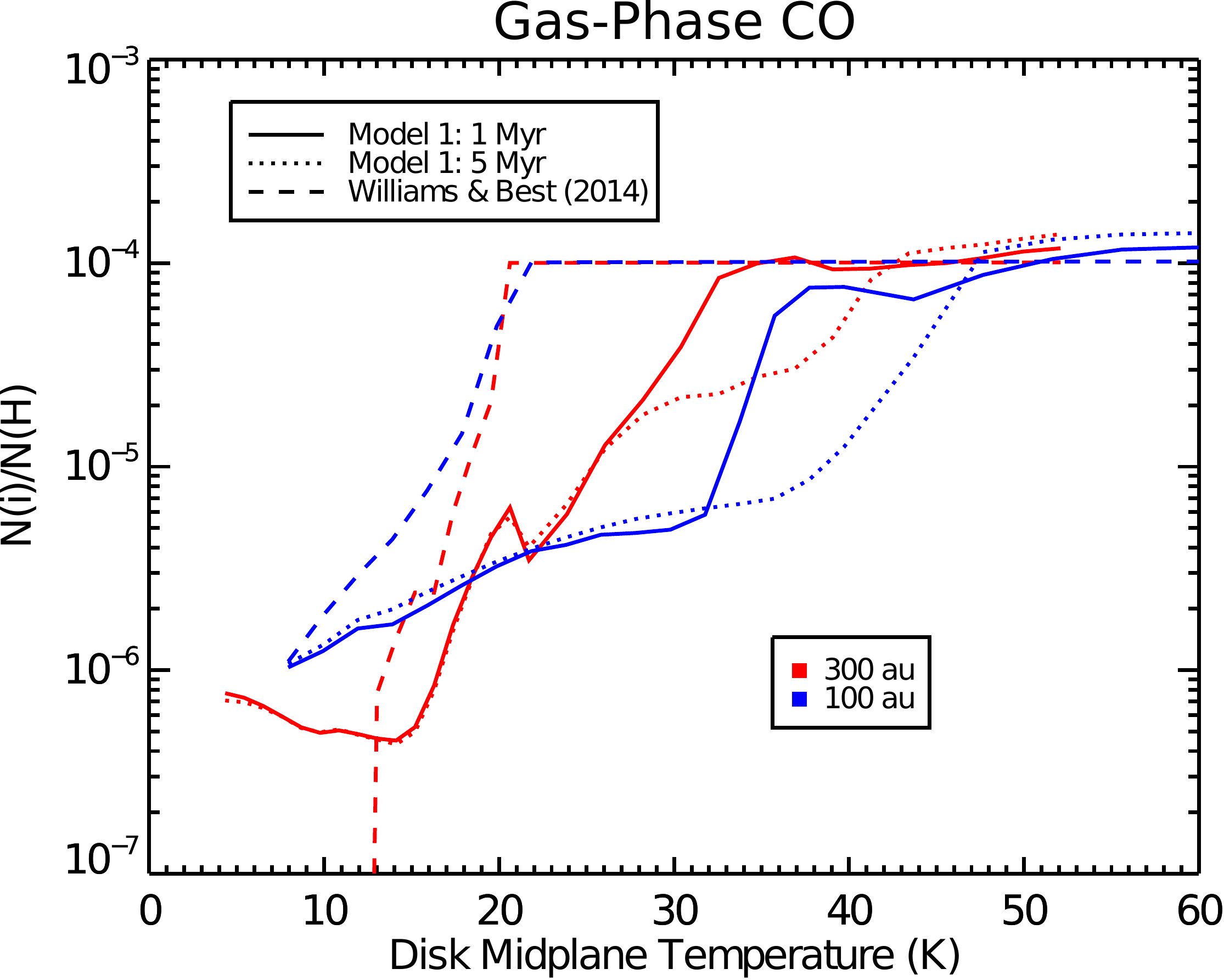}
  \caption{Average abundance of gas-phase CO as a function of the disk midplane temperature ($T_{\text{mid}}$). The results obtained with our Model 1 for disks of 1 and 5 Myr and at 100 and 300 au are compared to the model used by \citet{Williams&Best2014} for CO chemistry.}
\label{will}
\end{figure}

Our modeling reveals that, because of this processing on dust grains,
the CO chemistry does not behave in a simple way. Figure \ref{will} compares our average CO abundances with those predicted using the simple prediction of \citet{Williams&Best2014}, who assumed CO/H$_2$ = 0 where $T \leq 20$ K and in the photodissociation region, and CO/H$_2$ = 10$^{-4}$ elsewhere. Except for the coldest disks (where it predicts no CO), the simple prescription clearly overpredicts the average CO abundance for midplane temperatures below 30 to 35 K. At 20 K, the difference between our chemical model and the simple prescription is a factor of $\sim$ 40. This suggests that gas-phase CO abundance is critically dependent on the (dust) temperature distribution and grain surface chemistry. Figure \ref{will} shows that an abundance of  10$^{-4}$ is only reached for disks
with a midplane dust temperature above 30 K at 1 Myr and above 40 K at 5 Myr.

This complex CO chemistry may explain the low C$^{18}$O to dust ratio
observed in TW Hya by \citet{Favre2013}.
With an age estimate of 7-8 Myrs
\citep{Barrado2006,Ducourant2014}\footnote{a younger age of 3 Myr
has been quoted by \citet{Vacca&Sandell2011}, but seems inconsistent
with the dynamical age derived by \citet{Ducourant2014}}, TW Hya is
old enough compared to the timescale required to convert CO to
other forms on grain surfaces.
To estimate quantitatively the fraction of CO in gas phase,
a very accurate thermal model will be needed.
\citet{Hughes2008}
reproduced the mm continuum and CO images of the TW Hya disk
by two different disk models, a ``cold'' one with
$T(r) = 30 \text{K} (r/100 \mathrm{au})^{-0.5}$, and a ``warm'' solution with
$T(r) = 40 \text{K} (r/100 \mathrm{au})^{-0.2}$. We also note that \citet{Hughes2008}
assume $T_{\rm dust}$ = $T_{\rm gas}$, and the temperatures are
certainly biased towards somewhat
higher values because of the expected vertical temperature gradient
and of the difference between gas and dust temperatures. Nevertheless, for both cases, we
would predict abundances of order 10$^{-5}$, in reasonable agreement
with the results of \citet{Favre2013}. Thus, there may be no
need to invoke other mechanisms than the conversion of s-CO to
s-CO$_2$ which is the dominant process in our study.

When comparing for other objects, CO depletion due to surface chemistry
may still have difficulties in explaining
the apparent low CO to dust ratio derived by \citet{Chapillon2008}
for the very warm disk of CQ Tau, where temperatures may exceed 50 K.
However, \citet{Chapillon2008} only marginally resolved the source
and their quoted temperatures are for gas only. Higher resolution studies,
coupled to a more detailed thermal and chemical modeling for this source,
would be needed to conclude.

We did not consider the very strong UV flux of Herbig
Ae stars, so that a direct comparison between our results and any
specific object is premature. Similarly, we only
considered small grains in this study. Larger grains will reduce the
gas-grain interactions, thereby augmenting the conversion timescale
for CO, but as pointed out by \citet{Chapillon2008}, these large
grains could remain cooler.
Also, turbulent mixing may play a role in bringing more frequently
dust in the warm region, reducing the CO deficit. Turbulent mixing
has been studied  by \citet{Furuya&Aikawa2014}, and found to suppress
the carbon depletion only within the inner 50 au. High CO content is
thus expected in the inner regions (radius perhaps up to 30-50 au only).
This is consistent with the CO/H$_2$ ratio about 10$^{-4}$ derived
from IR measurements by \citet{France2014} for RW Aur.

\section{Conclusions}
In this paper, we have discussed the chemistry of carbon-bearing
species in protoplanetary disks using our gas-grain chemical model
Nautilus. In particular, we studied the effect of the disk temperature
structure on the chemistry of carbon reservoir in disks. We also
studied the impact of different initial parent cloud compositions.
We found that the reservoir of carbon in disks is not sensitive to
initial conditions such as gas density or C/O elemental ratio of the
parent molecular cloud.

However, carbon chemistry is strongly sensitive to the disk vertical
temperature profile. In many of our models, we have found that the CO
gas-phase abundance is lower than the canonical value of $\sim 10^{-4}$
generally assumed. Indeed, with our
models, such abundances are only reached for warm disks with a midplane
temperature above 30 K, at 100 and 300 au
for a disk age of 1 and 5 Myr. For smaller temperatures, we found that
CO is converted into other species at the surface of the grains, such
as s-CO$_{2}$, for which higher freezeout
temperatures prevent their desorption back into the gas-phase.
The effect is slightly less efficient at low cosmic-rays
ionization rate (down to 1.3$\times$10$^{-18}$ s$^{-1}$) and
large radii (300 au).

Finally, our work demonstrates that CO is not a simple tracer of H$_2$ in
protoplanetary disks, because its abundance is very sensitive
to the large (dust) temperature gradients that exists there. These gradients
will be very difficult to constrain with sufficient accuracy.
Given the additional complexity of isotopologue selective
photodissociation and fractionation, evaluating disk masses from
measurements of the CO isotopologue emission only will remain a very
challenging task.

\begin{acknowledgements}
The authors thank the French CNRS/INSU program PCMI for their partial
support of this work. VW and FH research is funded by the ERC Starting
Grant (3DICE, grant agreement 336474). The authors also thank the referee for suggestions which helped to improve this paper.
\end{acknowledgements}

\clearpage

\bibliography{biblio}
\bibliographystyle{aa}


\end{document}